\documentclass[11pt,a4paper]{article}
\usepackage{latexsym}
\usepackage{amsmath}
\usepackage{amsfonts}
\usepackage{amsbsy}
\usepackage{amssymb}
\usepackage{epsfig}
\usepackage{mathrsfs}
\usepackage{epsfig}
\usepackage{psfrag}
\usepackage{bbm}

\numberwithin{equation}{section}

\def\be{\begin{equation}}
\def\ee{\end{equation}}
\def\ba{\begin{eqnarray}}
\def\ea{\end{eqnarray}}

\newcommand\nn{\nonumber}
\newcommand\q{\quad}

\title{Perfect discretization of reparametrization invariant \\ path integrals}
\author{Benjamin Bahr$^{1,2}$, Bianca Dittrich$^2$, Sebastian Steinhaus$^2$\\[5pt]
\small $^1$ DAMTP, University of Cambridge,\\
\small  Wilberforce Road, Cambridge CB3 0WA, UK \\
\small   $^2$ MPI for Gravitational Physics,\\
 \small Am M\"uhlenberg 1, D-14476 Potsdam, Germany }

 \date{}

\begin{document}

\maketitle

\begin{abstract}
To obtain a well defined path integral one often employs discretizations. In the case of gravity and reparametrization invariant systems, the latter of which we consider here as a toy example, discretizations generically break diffeomorphism and reparametrization symmetry, respectively. This has severe implications, as these symmetries determine the dynamics of the corresponding system.

Indeed we will show that a discretized path integral with reparametrization invariance is necessarily also discretization independent and therefore uniquely determined by the corresponding continuum quantum mechanical propagator. We use this insight to develop an iterative method for constructing such a discretized path integral, akin to a Wilsonian RG flow. This allows us to address the problem of discretization ambiguities and of an anomaly--free path integral measure for such systems. The latter is needed to obtain a path integral, that can act as a projector onto the physical states, satisfying the quantum constraints.  We will comment on implications for discrete quantum gravity models, such as spin foams.

\end{abstract}


\section{Introduction}


Discretizations have become a popular tool in classical and quantum physics, on the one hand to allow for a numerical treatment, on the other hand to regularize for instance the path integral. Also a quite popular expectation is that fundamental physics is based on discrete structures rather than continuum space time, a conjecture followed in a number of quantum gravity approaches, for instance \cite{Dowker:2006wr, Konopka:2006hu}.

However, in discretizing a continuum theory one has to face several issues. One issue, which is particularly relevant for discrete quantum gravity approaches, is that the symmetries of the continuum theory might be broken by the discretization. Another issue is that discretizations are typically never unique, many different discrete models might lead to the same continuum physics.\footnote{Although one of the main problems in many discrete quantum gravity approaches is actually to extract large scale physics.} This is not a problem if discretization is just viewed as a tool for obtaining continuum physics, but has to be addressed if the discrete theory is claimed to be `fundamental'.

These problems appear in one form or other in many approaches to quantum gravity, where discretizations break diffeomorphism symmetry \cite{bahrdittrich1}. This symmetry is deeply entangled with the dynamics of the theory and hence its breaking has particularly severe repercussions \cite{bdreview}.

In the canonical framework diffeomorphism symmetry leads to the Hamiltonian and diffeomorphism constraints, satisfying the so--called Dirac algebra, a canonical version of the group property of diffeomorphisms. These constraints are central to the (quantum and classical) dynamics. A discretization, violating diffeomorphism symmetry leads however to a violation of the Dirac algebra and inconsistencies in the dynamics \cite{wp, lollreview}. Some of these issues can be addressed in the consistent and uniform discretization \cite{Gambini:2005vn} and master constraint \cite{Thiemann:2003zv} approach, however there the symmetries can only be regained in the continuum limit (if at all) \cite{gb2}.  There is an anomaly--free quantization of the Hamiltonian constraints \cite{Thiemann:1997rv} in Loop Quantum Gravity, however it features many ambiguities whose significance is not fully understood yet \cite{Perez:2005fn} and the status of the corresponding Dirac algebra is not fully satisfactory \cite{Thiemann:1997rv, Lewandowski:1997ba}.
Although one might suspect that in a  covariant approach the situation regarding diffeomorphism symmetry is much better than in the canonical ones, this is not the case -- indeed the problems are closely connected \cite{bahrdittrich1,bdreview,hartle, hohn}. So far, we are lacking a path integral in which diffeomorphism symmetry is realized on the discrete level. Related to this issue is the lack of `discretization independent'  (four--dimensional gravity) models, as there are arguments \cite{Pfeiffer:2003tx} that discretization independence should be the equivalent to the diffeomorphism invariance of the continuum. Indeed, we will present in section \ref{sec2} an argument involving the dynamics of the theory, showing this equivalence for the toy model of a one--dimensional reparametrization invariant system.

A proper (non--compact) gauge symmetry in a path integral leads to divergencies, as one integrates an invariant amplitude over gauge orbits. If the symmetries are broken, these divergencies do not necessarily appear on the discrete level, but should reappear in the continuum limit. However, these divergencies have  to be distinguished from proper UV divergencies which might also appear in a continuum limit.
Additionally in a perturbative approach to path integrals broken symmetries might lead to inconsistencies \cite{hohn}.

In addition to the amplitudes for the path integral one has to define an integration measure. Usually one would like this integration measure to be anomaly--free, i.e. to respect the gauge symmetries of the action. Although there is no discrete  amplitude respecting diffeomorphism invariance yet, there is some work \cite{Bojowald:2009im,Bianchi:2010fj,Bahr:2010my} deriving invariance conditions for the measure in spin foam models and in this way fixing some of the ambiguities. These however do not involve the full dynamics so far.  Again we will show in the context of the one--dimensional toy model that a symmetry preserving measure can be found. For this one has however to involve the full dynamics of the theory.\\[2pt]

There is a wide range of ideas to tackle these issues, which we cannot all discuss here. We will rather illustrate one approach, which is to obtain so--called perfect discretizations via a Wilsonian renormalization procedure \cite{Wilson:1973jj,Hasenfratz:1997ft,Bietenholz:1999kr,Bahr:2010cq}, within the toy model of a one--dimensional reparametrization invariant system.  The basic idea is to `pull back' continuum physics onto the lattice. Alternatively one can start with a discretization, subject it to a renormalization group transformation, which maps the physics of a fine discretization to a coarse grained one. Iterating this procedure one should be able to find a fixed point which gives the continuum physics exactly mirrored onto the discretization.

Although the model we will consider in this work is extremely simple compared to gravity, it displays many features that we believe will also be relevant for gravity. We will show that even for this simple system, a perfect discretization cannot be simply guessed, but has to be determined by solving the dynamics of the system. In particular we will show that a full implementation of reparametrization invariance -- the equivalent to diffeomorphism invariance in general relativity -- into the discrete path integral implies `discretization invariance', i.e. the result of the path integral is independent of the number of subdivisions, including the possibility to have none at all. We expect that a similar feature will hold for gravity. This `discretization invariance' actually means that the basic discrete amplitude (and measure) provides already the full quantum propagator of the system, also on large scale. Hence any free parameter in such a discretization will influence macroscopic physics. In this sense, requiring reparametrization invariance for the discretization, will resolve all discretization ambiguities. We can expect that likewise many ambiguities for discrete quantum gravity are fixed by requiring diffeomorphism invariance. The crucial question is, whether we are left with finitely or infinitely many (relevant) parameters \cite{Reuter:2007rv}.

As will be explained in section \ref{sec2} for the one--dimensional reparametrization invariant systems, a perfect discretization of the path integral is provided by the (continuum) quantum propagator, i.e. the evaluation of the continuum path integral. This is however not very helpful in the many situations where such a continuum quantum propagator or path integral is not available. In this case one typically rather has to start with the discrete system and develop some approximation methods. We therefore develop an iterative method to evaluate the path integral, basically following a Wilsonian renormalization approach. This approach will automatically address the issue of discretization ambiguities, as we do not consider a specific discretization but a  parameter space describing a certain class of discretizations, in which one determines the flow induced by a renormalization group transformation.

A technical difficulty which we address here, is to deal with the broken gauge symmetries in the path integral. This can be done by expanding around a special solution, for which the symmetries are exact. (Typically the zero solution $q_n\equiv 0$ for one--dimensional systems. For many discretizations of gravity flat space is such a solution.) The action (and measure) can then be improved order by order such that the gauge symmetries are exact to the order where one has to evaluate the path integral. Hence one can either gauge fix or cleanly separate the (infinite) factor arising from the integration over the gauge orbits.

To start with we will explain the basic ideas of the approach in the next section \ref{sec2}. In particular we will show that for one--dimensional systems reparametrization invariance and discretization independence for the discrete path integral are equivalent. In the following section \ref{Sec:q=0} we will consider the parametrized, discretized harmonic oscillator, and demonstrate the concepts from section \ref{sec2}.  Since in this case the time variable is discrete as well, this is not a linear system anymore, and we consider different linearizations in order to compute the reparametrization-invariant propagator. In section \ref{Sec:AHO} we will treat the (quartic) anharmonic oscillator with the same methods, computing the perfect propagator to first order in the interaction parameter $\lambda$. Finally, we will summarize and discuss our findings in section \ref{Sec:Summary}, considering in particular the implications for discrete gravity and spin foam models. The appendices \ref{app1}, \ref{unique}, \ref{App:BetaRecursion}  contain more technical details on the uniqueness of the solutions to the recursion relations describing the renormalization flow, and techniques to solve such recursion relations.

\section{Reparametrization invariance in the discrete and discretization independence}\label{sec2}

Consider a one-dimensional mechanical system defined by a Lagrangian $L(q,\dot q)$. The variation of the corresponding action $S=\int L(q,\dot q)$ with respect to $q(t)$ will determine the solutions. Such a system can be made reparametrization invariant by adding the time parameter $t$ as a dynamical variable to the configuration variable $q$. Time evolution is then with respect to an auxilary parameter $s$. One can define a new action (here $'$ denotes the derivative with respect to the auxiliary parameter $s$)
\ba\label{b01}
S\big(q(s),\,t(s)\big)\;&=&\;\int ds\;L\left(q,\frac{q'}{t'}\right)t'\
\ea
which is invariant under reparametrizations $q(s),t(s)\rightarrow q(f(s)),t(f(s))$ of the trajectories. This new action has to be varied with respect to $q(s)$ and $t(s)$. The equation of motion for $t(s)$ will however be automatically satisfied, if the equation of motion for $q(s)$ is. Hence our system is underdetermined as we have two variables but only one independent equation of motion. Indeed, the solutions $q_s(s)$, $t_s(s)$ are not uniquely determined by the boundary conditions, as given one such solution one can find a family of (physically equivalent) solutions $q_s(f(s)),t_s(f(s))$ by reparametrizing  the auxiliary evolution parameter $s\rightarrow f(s)$. Performing a (singular) Legendre transformation one will find instead of a proper Hamiltonian a Hamiltonian constraint $C=p_t+H$ (where $H$ is the Hamiltonian of the original system), which at the same time is the generator of the gauge transformations, i.e. it generates evolution in the auxiliary parameter $s$. This is very similar to general relativity (where  the role of the auxiliary parameter is taken over by space time coordinates).

This reparametrization symmetry is typically broken if we discretize the system. To this end we replace $q(s),t(s)$ with $s$ in some finite interval, by some finite set of variables $q_n,t_n$ with $n=0,\ldots,N$. One method of discretization is to replace derivatives by difference quotients and to choose some discretization for the potential term $V$, for instance
\ba\label{b02}
S(q_n,t_n)\;&:=&\;\sum _{n=0}^{N-1} S_n\;:=\;  \sum_{n=0}^{N-1}\left[\frac{1}{2}\left(\frac{q_{n+1}-q_n}{t_{n+1}-t_n}\right)^2
\,+\,\frac{1}{2}\left(  V(q_n)+ V(q_{n+1}) \right)\right](t_{n+1}-t_n)
\ea
(Here and in the following we will consider Wick rotated actions in order to make the path integrals convergent. This just changes the minus in front of the potential $V$ to a plus sign.) The factor $(t_{n+1}-t_n)$ arises from the integration measure $\int ds \, t'$ in (\ref{b01}).

This action (\ref{b02}) does in general (an exception being the free particle $V\equiv 0$) not feature anymore any gauge symmetries.  As a result, the equations of motion arising by varying with respect to $q_n$, $t_n$, $n=1,\ldots,N-1$ in fact uniquely fix both $q_n$ and $t_n$.  Hence both $q$ and $t$ become propagating (or physical) degrees of freedom in the discrete theory. Only in the continuum limit does reparametrization invariance arise, so that $t$ becomes a gauge variable again. The broken symmetries lead  to so-called pseudo constraints in the canonical formalism \cite{bahrdittrich1,Gambini:2005vn, hohn} instead of proper constraints. These are equations of motion, i.e. equations between the canonical data of two consecutive time steps $n$ and $n+1$,  which do however only weakly depend on the data at $n+1$. (Proper constraints are equations of motion which do involve only data of one time step $n$.)

This feature arises also for discretizations of gravity, such as Regge Calculus \cite{bahrdittrich1}. In this case, the discretization (i.e. the triangulation of space-time) breaks reparametrization-invariance (i.e. four-dimensional diffeomorphism invariance) as well. Furthermore, in gravity theories singling out an equivalent for $t$, i.e. identifying the (pseudo) gauge degrees of freedom and separating them from the truly physical ones, becomes hideously complicated.

The breaking of gauge symmetries is a result of the choice of discretization, however. In particular, there are discrete actions in $1D$ which exhibit a discrete remnant of the reparametrization invariance, which lead to the correct amount of physical degrees of freedom. Such \emph{perfect actions} can be constructed by a refinement process. Starting from a discrete action $S(q_n,t_n)$, one \emph{improves} it by refining the discretization, solving the equations of motion for the refined degrees of freedom, and evaluating the refined action on that solution. By iterating this process, or by directly considering the limit of sending the refined discretization to the continuum, one will find the perfect action, which can be  shown \cite{bahrdittrich1} to be given by Hamilton's principal function for the system, i.e.
\ba\label{b03}
S_{perf}(q_n,t_n)=\sum_n S_{HPF}(q_n,t_n,q_{n+1},t_{n+1})
\ea
where $S_{HPF}(q_i,t_i,q_f,t_f)$ is Hamilton's principal function (for the continuum system), i.e. the action evaluated on the solution $q(s)$, $t(s)$ with boundary conditions $(q_i,t_i,q_f,t_f)$. The perfect action does display a gauge symmetry for every (inner) discretization point $n$, which we will call vertex translation symmetry, as it is of the form (for finite gauge parameter $\lambda$)
\ba\label{b04}
t_n &\rightarrow& t_n+\lambda \nn\\
q_n  &\rightarrow& f(q_{n-1},t_{n-1},q_n,t_n,q_{n+1},t_{n+1}) \q ,
\ea
i.e. it translates the discretization point in time. Hence all the $t_n$ can be seen as gauge parameters. Furthermore the action is `discretization independent', i.e. Hamilton's principal function $S_{HPF}(q_0,t_0,q_N,t_N)$ (being the classical equivalent to the partition function or path integral, which encodes all the dynamical information) computed from $S_{perf}(q_n,t_n)$ does not depend on the number of discretization points used, and coincides with the one computed from the continuum action $S_{HPF}(q_0,t_0,q_N,t_N)=S_{perf}(q_0,t_0,q_N,t_N)$. In particular we can choose $N=1$. Note that to construct this action one needs to basically solve the dynamics of the system. This can be seen as a huge disadvantage of the method. However Hamilton's principal function provides actually the only discretization featuring vertex translation symmetry \cite{marsdenwest}.  Hence addressing the dynamics of the system cannot be avoided if one wants to implement the symmetries of the system (or even just ensure that these symmetries appear in the continuum limit).

Instead of solving the dynamics at once, one can consider various approximations and follow an iterative approach of constructing   better and better actions. We will follow this idea here and see that it is closely related to a Wilsonian renormalization group approach. The perfect actions then arise as fixed points of iterative  (renormalization group) transformations. Note that this also resolves (at least in one dimension) all the discretization ambiguities, as $S_{perf}$ is the unique action, displaying (the discrete remnant) of discretization invariance.

It has been shown that this procedure also works for $3D$ Regge Calculus with nonzero cosmological constant \cite{bahrdittrich2}. For first steps to dealing with $4D$ gravity in a perturbative set--up, see \cite{Bahr:2010cq}.

This addresses the classical theory. An open issue is, whether a similar approach will work for the quantum theory, here in a path integral approach. To evaluate path integrals analytically we have to follow a perturbative approach. For broken symmetries such a perturbative ansatz may turn out to be inconsistent however \cite{hohn}. The problem are exceptional solutions (such as $q_n\equiv 0$), which are typically the only ones displaying symmetry under vertex translations (i.e. the $t_n$ can be chosen arbitrarily, the $q_n$ remain zero). Perturbing around such solutions one will not find a quadratic term for the time variables, however these time variables will appear in the higher order potential terms. This hinders the perturbative evaluation of the path integral.  A way out, suggested in \cite{hohn,Bahr:2010cq} and also followed up here, is to improve the action perturbatively, so that gauge invariance can be obtained order by order. This allows an evaluation of the path integral to the corresponding order, either by gauge fixing or by changing to gauge invariant variables.

In addition to the action, a path integral requires an integration measure. In the presence of gauge symmetries one would require this measure to be invariant under these symmetries, otherwise one will obtain anomalies. Indeed only if the measure is invariant, can the path integral serve as a projector onto the states satisfying the constraints (arising in a canonical quantization) \cite{hartle}. We will argue here that, similarly to having to solve the classical dynamics to obtain the perfect action, one needs to solve the quantum dynamics to obtain the `perfect measure'\footnote{The split of the path integral into amplitude and measure is ambiguous and we will not insist on one particular splitting here.}  and with this a perfect discretization of the path integral:

Consider a discrete path integral with two time steps
\ba\label{b05}
\langle q_0,t_0|q_2,t_2\rangle:=Z(q_0,t_0,q_2,t_2)\;:=\;\int dq_1 dt_1\;K(q_0,t_0,q_1,t_1)K(q_1,t_1,q_2,t_2)
\ea
where we summarised amplitude and measure for one discretization step into the discrete propagator $K(q_n,t_n; q_{n+1},t_{n+1})$. (We term this the propagator as the perfect discretization is given by the quantum propagator of the system.) Now assume that vertex translation invariance (\ref{b04})  has been fully implemented into (\ref{b05}). As the gauge symmetry just translates the time variable $t_1$ we can gauge fix to an arbitrary value $t_1=t_1^f$ and drop the $t_1$ integration. (The Fadeev-Popov determinant is trivial in this case.) We obtain
\ba\label{b06}
Z(q_0,t_0,q_2,t_2)\;:=\;\int dq_1 \;K(q_0,t_0,q_1,t_1^f) K(q_1,t_1^f,q_2,t_2) \q .
\ea
As $t^f_1$ is arbitrary we can consider the limit $t^f_1 \rightarrow t_2$.  Now the interpretation of $K(q_1,t_1^f,q_2,t_2)$ is to give the amplitude for a particle to propagate from $q_1$ to $q_2$ during the time interval $(t_2-t_1^f)$. In the limit $t^f_1 \rightarrow t_2$ we should therefore have $K(q_1,t_1^f,q_2,t_2)\rightarrow \delta(q_1-q_2)$. Hence we will obtain that the right hand side of (\ref{b06}) is equal to the discrete propagator $K(q_0,t_0,q_2,t_2)$ and since $t_1^f$ can be chosen arbitrarily we have shown that
\ba\label{b07}
K(q_0,t_0,q_2,t_2)\;=\;\int dq_1 \;K(q_0,t_0,q_1,t_1) K(q_1,t_1,q_2,t_2) \q .
\ea
Starting from the assumption, that vertex translation invariance has been realized for the path integral (\ref{b05}), we have shown that the discrete propagator $K$ needs to satisfy (\ref{b07}), which is the usual convolution property for the propagator kernel in quantum mechanics.

This actually proves that a path integral with vertex translation invariance is also discretization invariant, i.e. it does not depend on the number of subdivisions. Having no subdivisions at all -- which gives just the discrete propagator $K$ -- should coincide with having infinitely many -- which gives the propagator in the continuum. Hence the discrete propagator $K(q_n,t_n,q_{n+1},t_{n+1})$ is given by the usual quantum mechanical propagator for the continuum system.

On the other hand, if we have a discrete propagator satisfying (\ref{b07}) (where we do not integrate over $t$), then the corresponding path integral is invariant under vertex translations, as one can integrate out and reinsert every discretization point $(q_n,t_n)$. This shows that for one--dimensional reparametrization invariant systems, finding a discretization which respects this invariance and discretization independence\footnote{Where here we understand under discretization independence the property (\ref{b07}). There we do not integrate over the time variable, that is we consider already a gauge fixed version of the path integral.  Otherwise the integral would be divergent, if vertex translation invariance is realized.}  are equivalent. We conjecture that this will also hold for discrete gravity, i.e. discretized  path integrals respecting diffeomorphism invariance (which in its discrete form is also expected to be vertex translation invariance, see the discussion in \cite{Bahr:2010cq}) should be also discretization independent. Again, this also means that in order to construct such path integrals one has to consider the dynamics of the discrete (quantum) models.

To summarize the discussion, a perfect discretization of the quantum mechanical path integral would be given by the (continuum) propagator. This would however not be very helpful for systems in which this propagator is not known, such as quantum gravity. Therefore in this article, we want to adopt the procedure of successively improving the action of a classical discrete system to the quantized case, as this approach might also be helpful for more complicated cases. This will also introduce a method to actually solve the path integral for the corresponding continuum system iteratively.

To improve the discrete propagator iteratively we start from the propagator property (\ref{b07}) with a 'naively discretized' propagator $K^{(0)}$. This can be taken as
\ba\label{b08}
K^{(0)} (q_n,t_n,q_{n+1},t_{n+1})=\eta^{(0)}_n \exp(-\frac{1}{\hbar} S^{(0)}_n) \q
\ea
where $S^{(0)}$ is the naive discretized action (\ref{b02}). We will see that the classical part of the iteration equations (which is the part without $\hbar$ dependence) leads to the perfect discretization for the action. The initial measure factor $\eta^{(0)}_n$ should be chosen such that $K^{(0)}\rightarrow \delta(q_n-q_{n-1})$ for $(t_{n-1}-t_n)\rightarrow 0$.

Refining the discretization and integrating out the intermediate degrees of freedom will result in the improved propagator $K^{(1)}$, and iterating the procedure will lead to the perfect propagator satisfying (\ref{b07}). From the construction it should be clear that one subdivision step will be sufficient, i.e. we define
\begin{eqnarray}\label{Gl:PropagatorRecursion}
K^{(n+1)}(q_0,t_0,q_2,t_2)\;:=\;\int dq_1 dt_1\;K^{(n)}(q_0,t_0,q_1,t_1)K^{(n)}(q_1,t_1,q_2,t_2)  \q .
\end{eqnarray}

Here we included the integration over $t_1$. In section \ref{Sec:q=0}  we will perturb around the solution $q_n\equiv 0$ which is reparametrization invariant also in the naive discretization and introduce a method such that the $t_1$ integration can be dropped.

\vspace{0.5cm}
\begin{figure}[hbt!]
\begin{center}
    \psfrag{x0}{$q_0,t_0$}
    \psfrag{x1}{$q_1,t_1$}
    \psfrag{x2}{$q_2,t_2$}
    \psfrag{K0}{$K^{(n)}(q_0,t_0,q_1,t_1)$}
    \psfrag{K1}{$K^{(n)}(q_1,t_1,q_2,t_2)$}
    \psfrag{K2}{$K^{(n+1)}(q_0,t_0,q_2,t_2)$}
    \psfrag{T1}{Integrate out $q_1$, $t_1$.}
    \psfrag{T2}{$K^{(n)}\rightarrow K^{(n+1)}$}
    \includegraphics[scale=0.9]{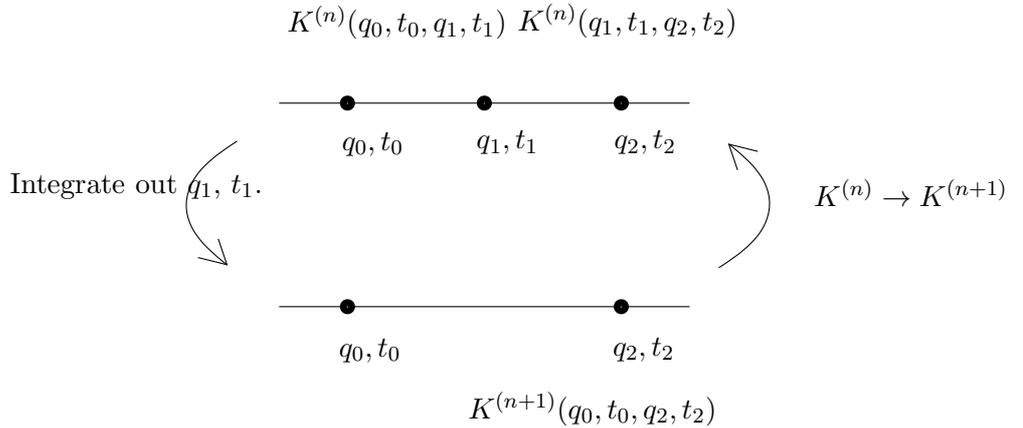}
    \caption{The refinement process of the propagator $K^{(n)}$ involves subdividing the discretization intervals and integrating over the new variables, obtaining a new propagator $K^{(n+1)}$. This process can be iterated, leading in the limit to the perfect propagator, which is in particular invariant under refinement of the discretization.}\label{Fig:Coarse}
\end{center}
\end{figure}

In order to find the perfect propagator, we will parameterize (some part of) the space of functions, allowing for arbitrary couplings in the action and measure factors. Equation (\ref{Gl:PropagatorRecursion}) will then lead to recursion relations for the coefficients, which are closely related to the renormalization group flow of the system. We will find that, not only does the classical perfect action provide a fixed point of the equations derived from (\ref{Gl:PropagatorRecursion}), but also the measure factor $\eta$ will be determined, providing the quantum corrections to the classical perfect action.

\section{The harmonic oscillator linearized around $\bar q_n=0$}\label{Sec:q=0}

In the following, we will deal with the (Wick-rotated) parameterized harmonic oscillator, the action for which is given by
\begin{eqnarray}\label{Gl:ActionParameterizedContinuous}
S\big(q(s),\,t(s)\big)\;&=&\;\frac{1}{2}\int ds\;\left(\frac{q'(s)^2}{t'(s)}+\omega^2q(s)^2t'(s)\right) \q .
\end{eqnarray}

By naively discretizing the action one replaces continuum functions $q(s)$, $t(s)$ by discretely many values $q_n=q(s_n)$, $t_n=t(s_n)$, and derivatives by difference quotients. For the discretization of the integral measure we use $ds=s_{n+1}-s_{n}$. Hence we arrive at
\begin{eqnarray}\label{Gl:DiscretizedParametrizedHO}
S^{(0)}(q_n,t_n)\;&=&\;\sum_{n=0}^{N-1}\left[\frac{1}{2}\left(\frac{q_{n+1}-q_n}{t_{n+1}-t_n}\right)^2
\,+\,\frac{\omega^2}{2}\left(\frac{q_n^2+q^2_{n+1}}{2}\right)\right](t_{n+1}-t_n) \q .
\end{eqnarray}
As we have already indicated, because of the naive discretization, the action (\ref{Gl:DiscretizedParametrizedHO}) does not contain  gauge symmetries, see for instance \cite{bahrdittrich1}. That is given (general) boundary data $(q_0,t_0;q_N,t_N)$ both the $q_n$ and the $t_n$ for $n=1,\ldots N-1$ are determined by the equations of motion.

The action (\ref{Gl:DiscretizedParametrizedHO}) is \emph{not} quadratic and not even polynomial in the $t_n$-variables, and therefore not considered to be a non--interacting system. The path integral is therefore non-trivial, and we consider approximations to it which will render it feasible. The approximation we are going to perform here involves quantizing only the perturbations around a background solution, i.e. we consider the perturbed variables
\begin{eqnarray}\label{Gl:Linearize}
q_n\;&=&\;\bar q_n\,+\,x_n\\[5pt]
t_{n}\;&=&\;\bar t_n\,+\,\tau_n
\end{eqnarray}

\noindent where $\bar q_n$, $\bar t_n$ are to satisfy the equations of motion coming from the discrete action. The path integral for this linearized theory, containing the variables $x_n$, $\tau_n$, can be viewed as a linear approximation to the path integral for the whole theory. Of all the solutions to the equations of motion derived from (\ref{Gl:DiscretizedParametrizedHO}), the constant solution, i.e. $\bar q_n\equiv 0$, $\bar t_n$ arbitrary, is special in the regards that this solution still exhibits some gauge symmetry (i.e. change in the $\bar t_n$), which is a remnant of the reparametrization invariance of the continuum theory. In this regards, it is analogous to the flat solution $\epsilon_t=0$ in $4D$ Regge calculus \cite{bahrdittrich1, Rocek}, which still exhibits the vertex displacement symmetry. We will consider linearization around this solution first, before turning to linearization around a general solution later.

Expanding the action for one time-step (neglecting all orders of $x_n$, $\tau_n$ higher than two), we arrive at
\begin{eqnarray}\label{Gl:HOLinearizedAroundq=0}
S^{(0)}(x_0,\tau_0,x_1,\tau_1)\;&=&\;\frac{1}{2}\frac{(x_{1}-x_0)^2}{T_{01}}\,+\,\frac{\omega^2}{4}(x_1^2+x_0^2) T_{01}\\[5pt]\nonumber
&=:&\;\alpha_1^{(0)}(T_{01})(x_0^2+x_1^2)\,+\,\alpha_2^{(0)}(T_{01})x_0x_1
\end{eqnarray}

\noindent with $T_{01}:=\bar t_1-\bar t_0$. Note that here the $\bar t_n$ (hence the $T_{nm}$) now play the role of parameters, since they are just  the background solution. The actual perturbed time variables $\tau_n$ have disappeared from (\ref{Gl:HOLinearizedAroundq=0}), mirroring the invariance of the background solution $\bar q_n=0$ under translations of the  $\bar t_n$.

In the second line of (\ref{Gl:HOLinearizedAroundq=0}) we introduced a general parametrization of an action, quadratic in the variables $x_0,x_1$ and symmetric under exchanging $x_0,x_1$. This allows also to consider different discretizations in one go. We will define a renormalization procedure (corresponding to the so-called decimation procedure, where one just integrates out some variables) on the space of such actions and look for  fixed points of the renormalization flow acting on the parameter functions $\alpha_k$. Hence we can expect that some choices for the initial discretization are not relevant, as these different choices may flow to the same fixed point.

A difference to the usual discussions of renormalization group transformations is, that the parameters $\alpha_k(T_{01})$ in the action (\ref{Gl:HOLinearizedAroundq=0}) are not just coupling constants but functions of the time distance $T_{01}$. The appearance of `coupling functions'  $\alpha(T_{01})$ can be avoided by expanding these functions in a suitable basis, for instance a power series in $T_{01}$. In general this will introduce infinitely many coupling parameters (which one could also treat perturbatively). However the fixed point conditions (which are related to the condition of discretization independence and reparametrization invariance) will fix almost all of these coupling constants.

To define the iteration procedure we assume that the one--step propagator has the form
\ba
K(x_0,x_1,T_{01})&=& \eta(T_{01}) \exp\left[-\frac{1}{\hbar}\left(    \alpha_1(T_{01})(x_0^2+x_1^2) \,+\, \alpha_2(T_{01})\,x_0x_1   \right) \right] \q ,
\ea
that is that the measure factor $\eta(T_{01})$ does only depend on the background time difference and not for instance on the $x_0,x_1$. This assumption is justified, as the form of the propagator will be stable under iteration.

Since the perturbative variable $\tau_1$ does not appear in the action (\ref{Gl:HOLinearizedAroundq=0}) the propagator for two time steps
\begin{eqnarray}\label{b1}
K^{(1)}(x_0,x_2,T_{01}+T_{12})&=&\int dx_1d\tau_1\,\,\eta^{(0)}(T_{01})   \eta^{(0)}(T_{02}) \nn\\
 && \q\q\q  \exp\left[-\frac{1}{\hbar}\left(S^{(0)}(x_0,\tau_0,x_1,\tau_1)+S^{(0)}(x_1,\tau_1,x_2,\tau_2)\right)\right]\q\q
\end{eqnarray}

\noindent leads to a divergent result.

This infinite factor represents the volume of the discrete diffeomorphism group (in this case the actual placement of the interior discretisation point), which is the whole real line. We will drop this infinite factor here, so we will not integrate over $\tau_1$, which will lead to a finite result.\footnote{This is reminiscent of the Ponzano-Regge model, which also needs to be gauge fixed in order to produce finite results. One could argue that $\tau_1$ shouldn't range over the whole real line, since $t_1$ should always remain between $t_0$ and $t_2$. Indeed, restricting the integration range appropriately would lead to a finite result, just as a similar restriction within the Ponzano Regge model -- which ensures that one only sums over diffeomorphisms which preserve the orientation everywhere -- renders the model finite. However, the finite result does not only not agree with the Ponzano-Regge amplitude, it also leads to a triangulation-dependent model, in which the diffeomorphism symmetry is still broken \cite{PR}. We therefore do not adapt that strategy here.}

Having dropped the $\tau_1$-integration in the path  integral, the remainder can be easily computed. Choosing $T:=T_{01}=T_{12}$ for simplicity\footnote{It will turn out that this regular subdivision is sufficient to regain the full symmetry under vertex translations and arbitrary subdivisions, i.e. discretization independence.}, we obtain for the propagator
\begin{eqnarray}\nonumber
K^{(n+1)}(x_0,,x_2,2T)&=&\eta^{(n)}(T)^2\int dx_1\exp\left[-\frac{1}{\hbar}\left(
\alpha_1^{(n)}(T)(x_0^2+2 x_1^2 + x_2^2) + \alpha_2^{(n)}(T)(x_0x_1+x_1x_2)
\right)\right]\\[5pt]
&=:&\eta^{(n+1)}(2T)\exp\left[-\frac{1}{\hbar}\left(
\alpha_1^{(n+1)}(2T)\,(x_0^2+ x_2^2) + \alpha_2^{(n+1)}(2T) \,x_0x_2
\right)\right] \q .
\end{eqnarray}

\noindent This  defines recursion relations for the coefficients $\alpha_1$, $\alpha_2$ and $\eta$ given by
\begin{eqnarray}\label{Gl:HORec1}
\alpha_1^{(n+1)}(2T)\;&=&\;\alpha_1^{(n)}(T) - \frac{\alpha_2^{(n)}(T)^2}{8\alpha_1^{(n)}(T)}\\[5pt]\label{Gl:HORec2}
\alpha_2^{(n+1)}(2T)\;&=&\;-\frac{\alpha_2^{(n)}(T)^2}{4\alpha_1^{(n)}(T)}\\[5pt]\label{Gl:HORec3}
%
%
\eta^{(n+1)}(2T)\;&=&\;\sqrt{\frac{\pi\hbar}{2\alpha_1^{(n)}(T)}}\,\,\eta^{(n)}(T)^2 \q .
\end{eqnarray}

\noindent To find the perfect propagator, one can iterate the equations (\ref{Gl:HORec1}) - (\ref{Gl:HORec3}), with initial values for the $\alpha_i^{(0)}(T)$ taken from (\ref{Gl:HOLinearizedAroundq=0}). Alternatively we can directly look for the fixed points. Considering  the first two equations involving only the $\alpha$ coefficients, a family of fixed points is given by\footnote{The fixed point property is easy to check using $\sinh(2y)=2\sinh(y)\cosh(y)$ and  $\cosh(2y)=\cosh^2(y)+\sinh^2(y)$.}
\ba\label{b2}
S^*(x_0,x_1,T)&:=&\alpha_1^*(T)(x_0^2+x_1^2)+\alpha_2^*(T)x_0x_1 \nn\\
&=& \frac{\tilde \omega}{2g} \frac{\cosh(\tilde \omega T)(x_0^2+x_1^2)-2x_0x_1}{\sinh(\tilde \omega T)}  \q .
\ea
Note that this action is Hamilton's principal function for the harmonic oscillator with frequency $\tilde \omega$ and a coupling $1/g$ in front of the action. These constants are determined by the initial values for the action, which for our choice (\ref{Gl:HOLinearizedAroundq=0}) leads to $\tilde \omega=\omega,\,g=1$. Indeed \cite{SebDipl}, this initial action converges to (\ref{b2}) under the iterations defined by (\ref{Gl:HORec1}). That there is at least a two--parameter family of fixed points parametrized by  $\tilde\omega,g$ can be easily deduced from the iteration equations (\ref{Gl:HORec1}). These are invariant under a rescaling $\alpha_k \rightarrow g^{-1} \alpha_k$. Furthermore if $\alpha_k(\cdot)$ is a fixed point, so is $\alpha_k(\tilde \omega \times \cdot)$. In the appendix \ref{unique} we will show that the action (\ref{b2}) provides the most general solution to the fixed point conditions. 

For the fixed point equation of the measure factor $\eta(T)$ we can use the fixed point solution $\alpha_1^*(T)$. A solution is given by
\ba\label{etas}
\eta^*(T)=\sqrt{\frac{\tilde\omega}{2\pi \hbar g \sinh(\tilde \omega T)}} \,\, \exp( \tilde \xi \, T)  \q .
\ea
where $\tilde \xi$ is a free parameter. Again we will show in the appendix \ref{unique} that this is the most general fixed point solution. The free parameter  $\tilde \xi$ describes the possibility to add a  constant potential term $\sim T$ to the action (\ref{Gl:HOLinearizedAroundq=0}).

Note that the iteration equation for the measure factor $\eta$ is not linear in $\eta$ -- rather we have $\eta$ quadratically appearing on the RHS of (\ref{Gl:HORec3}). Hence if we scale the initial value $\eta^{(0)}$ with a factor $a$ we obtain a scaling $\eta^{(n)}\sim a^{2^n}$ for the $n$-th iteration. Therefore many initial values will either diverge or converge to zero. Starting with e.g. the measure factor of the path integral for the free particle  $\eta^{(0)}(T) = \sqrt{\frac{1}{2 \pi g \hbar}} \, T^{-\frac{1}{2}}$ leads to a non--vanishing convergent result however \cite{SebDipl}. This also satisfies the normalization condition mentioned in section \ref{sec2}. \\[5pt]

To summarize for the initial action (\ref{Gl:HOLinearizedAroundq=0}) we obtain the perfect quantum propagator
\begin{eqnarray}\label{Gl:PerfectPropagator}
K(x_0,x_1,T)\;&=&\;\sqrt\frac{\omega}{2\pi\hbar\sinh(\omega T)}\exp\left[-\frac{\omega}{\hbar}\frac{\cosh(\omega T)(x_0^2+x_1^2)-2x_0x_1}{2\sinh(\omega T)}\right]  \q .
\end{eqnarray}

Note that this propagator (\ref{Gl:PerfectPropagator})  -- although calculated for the linearized theory -- coincides with the non--perturbative quantum mechanical propagator for the harmonic oscillator
\ba\label{bc1}
K(q_0,t_0,q_1,t_1)&=&\sqrt\frac{\omega}{2\pi\hbar\sinh(\omega (t_1-t_0))}\exp\left[-\frac{\omega}{\hbar}\frac{\cosh(\omega (t_1-t_0))(q_0^2+q_1^2)-2q_0q_1}{2\sinh(\omega (t_1-t_0))}\right]  \q .\q\q
 \ea
 This  also gives the propagator for the parametrized harmonic oscillator -- which we obtain from our calculation for the linearized theory by replacing the background parameters $\overline{t}_n$ by the full dynamical variables $t_n$ and the perturbations $x_n$ by $q_n$.

Hence the propagator property
\ba\label{bc1a}
K(q_0,t_0,q_2,t_2)&=&\int dq_1 \,\, K(q_0,t_0,q_1,t_1)\,\,K(q_1,t_1,q_2,t_2)
\ea
is not only satisfied for equal step times $(t_2-t_1)=(t_1-t_0)$ (which holds due to the fixed point condition) but also for arbitrary subdivisions into unequal time steps.\footnote{This is not so surprising as we basically computed the path integral iteratively. The fixed points correspond to taking the (equal time step) discretization to the continuum limit.} Furthermore we achieved full `discretization independence': if we use for the discrete propagator in the (discretized) path integral the perfect propagator the result will not depend on the number of subdivisions we use.

As explained in section \ref{sec2}, related to this discretization independence is the invariance of  the path integral (\ref{bc1a}) under `vertex translations', i.e.~changing the value for $t_1$ in (\ref{bc1a})\footnote{The gauge symmetry also involves a corresponding change in $q_1$, this can however be absorbed in a variable transformation for $q_1$ in the action. One can check that the Jacobian of this transformation changes the measure factor accordingly, so that  one obtains the measure appropriate for the transformed $t_1$.}. Again, we do not integrate over $t_1$ (which in the parametrized theory is a dynamical variable) in (\ref{bc1a}), as this would just lead to an infinite factor.

The gauge symmetry in the path integral leads to a constraint in the canonical theory. Indeed,
the propagator (\ref{bc1}) satisfies the quantum mechanical constraint equation for the parametrized harmonic oscillator (which is just the usual Wick rotated Schr\"odinger equation if one reinterprets $t$ as a non--dynamical time parameter)
\ba\label{c1}
\hat{C}K(q_0,t_0,q_1,t_1)= \Big(\hbar \frac{\partial}{\partial t_0} + \frac{\hbar^2}{2} \frac{\partial^2}{\partial q_0^2} + \frac{\omega^2}{2} q_0^2\Big) K(q_0,t_0,q_1,t_1) =0  \q .
\ea

Note that this constraint equation is only satisfied for the fixed point (\ref{bc1}) and not for any of the propagators $K^{(n)}$ for finite $n$, which would correspond to some (non--perfect) discretized version of the path integral. That is the fixed point is characterized by  a symmetry, which is reflected in the constraint equation (\ref{c1}).

\subsection{Linearized dynamics around a solution with $\bar q_n\neq 0$} \label{secqq}

In this section we will linearize the discretised, parametrised harmonic oscillator around a classical solution (\ref{Gl:Linearize}) with $\bar q_n\neq 0$.  Since in this case the background solution does not exhibit gauge symmetry, neither does the linearized theory \cite{hohn}. Hence, the $\tau_n$ appear as variables and have to be dealt with.

Expanding the action (\ref{Gl:DiscretizedParametrizedHO}) into second order via (\ref{Gl:Linearize}), one arrives at an action for the linearized variables $x_n$, $\tau_n$. Since reparametrization symmetry is broken, there is no obvious choice for a gauge variable (a role which was fulfilled by the $\tau_n$ in the previous case, where the symmetries were present), so in the path integral the integrals over both $x_n$ and $\tau_n$ have to be performed.  Moreover, the resulting propagator is \emph{not} a projector on the physical Hilbert space of the continuum theory. 
\\

Performing one refinement step therefore amounts to evaluating
\begin{eqnarray}\label{Gl:RecursionLinearizedQNeq0}
K^{(n+1)}(x_0,\tau_0,x_2,\tau_2)\;=\;\int dx_1d\tau_1 K^{(n)}(x_0,\tau_0,x_1,\tau_1)K^{(n)}(x_1,\tau_1,x_2,\tau_2)
\end{eqnarray}

\noindent with $K^{(0)}(x_0,\tau_0,x_1,\tau_1)=\eta^{(0)}\exp(-S^{(0)}(x_0,\tau_0,x_1,\tau_1)/\hbar)$. In this case $\eta^{(n)}$ and the action $S^{(n)}$ depend furthermore on the background variables $\bar q_i,\bar t_i$. It should be noted that, for one refinement step, the background solution $\bar q_1$, $\bar t_1$ has to be chosen such that it is a solution to the equations of motion of the discretized action $S^{(n)}$ with boundary values $\bar q_0$, $\bar t_0$, $\bar q_2$, $\bar t_2$.

The integral (\ref{Gl:RecursionLinearizedQNeq0}) is actually finite in most cases\footnote{Note however that even in the Wick-rotated framework the path integral for the linearized theory (\ref{Gl:RecursionLinearizedQNeq0}) is not necessarily always convergent. The reason is that, depending on the (background) boundary values $q_0$, $t_0$, $q_N$, $t_N$, the Hessian matrix of second derivatives, which arises as inverse propagator for the $x_n$, $\tau_n$ due to the expansion (\ref{Gl:Linearize}), might not be positive definite, rendering the Gaussian integral divergent. This is also an effect of the broken gauge symmetries.
There is, however, a good control over which boundary values lead to positive definite Hessian \cite{SebDipl}, and we assume that such a choice has been made in what follows.}, despite the $\tau_1$ integration actually being carried out. This is a result of the broken symmetries of the linearized theory: the $\tau_n$ are actually propagating degrees of freedom, corresponding to a small but non--vanishing eigenvalue in the Hessian of the action.

Due to $\tau_1$ being present, there are many more coefficients in the action $S^{(n)}$, leading to more complicated recursion relations coming from (\ref{Gl:RecursionLinearizedQNeq0}). These have been derived in \cite{SebDipl}, and we report on the findings in what follows:

The recursion relations for the coefficients coming from the action $S^{(n)}$ converge to a fixed point which is given by
\begin{eqnarray}\label{fp1}
S\;&=&\;\frac{\omega}{2}\frac{\cosh(\omega T)(\bar q_0^2+\bar q_1^2)-2\bar q_0\bar q_1}{\sinh(\omega T)}\,+\frac{\omega}{2\sinh(\omega T)}\big(\cosh(\omega T) (x_0^2+x_1^2)-2x_0x_1\big)\\[5pt]\nonumber
& &\quad+\;\frac{\omega^2(\bar q_0-\cosh(\omega T)\bar q_1)}{\sinh^2(\omega T)}x_0(\tau_0-\tau_1)+\frac{\omega^2(\bar q_1-\cosh(\omega T)\bar q_0)}{\sinh^2(\omega T)}x_1(\tau_0-\tau_1)\\[5pt]\label{Gl:PerfectActionQNeq0}
& & \quad -\;\frac{\omega^4(\bar q_0-\cosh(\omega T)\bar q_1)(\bar q_1-\cosh(\omega T)\bar q_0)}{2\sinh^4(\omega T)}(\tau_0^2+\tau_1^2)
\end{eqnarray}

\noindent with $T=\bar t_1-\bar t_0$. Not surprisingly, the action (\ref{Gl:PerfectActionQNeq0}) is actually the perfect action of the harmonic oscillator expanded up to second order in the linearized variables (\ref{Gl:Linearize}). More interesting is the behavior of the measure factor $\eta^{(n)}$. One can separate the measure factor into two contributions
\begin{eqnarray}
\eta^{(n)}\;=\;A^{(n)}B^{(n)}
\end{eqnarray}

\noindent where $A^{(n)}$ arises as prefactor from the integration over the $x_1$, and $B^{(n)}$ from the integration over $\tau_1$ in (\ref{Gl:RecursionLinearizedQNeq0}). It is not hard to show that in the limit of infinite refinement
\begin{eqnarray}
A^{(n)}\;\longrightarrow\;\sqrt\frac{\omega}{2\pi\hbar\sinh(\omega T)}
\end{eqnarray}

\noindent which coincides with the measure factor for the linearized case (\ref{Gl:PerfectPropagator}). However, despite $B^{(n)}$ being finite for every $n$, in the limit it actually diverges, leading to a diverging propagator.

This is also not surprising if one keeps in mind that in the continuum theory $\tau$ is actually the gauge variable, and therefore should not be integrated over in the path integral. The breaking of symmetry in the discrete theory has led to a finite integral, however in the perfect limit the integration over $\tau_1$ coincides with an integral over the volume of the gauge group. This volume is infinite, since it corresponds to the placement of the intermediate point $\tau_n$, which can be put everywhere on the real line.

Hence, in the perfect limit, the gauge symmetry can also be recognized by the volume of the gauge group still being present in the path integral. This could have been remedied from the first place by not integrating over $\tau_1$, but rather gauge fixing it to a definite value, e.g. $\tau_1=(\tau_0+\tau_2)/2$, which also coincides with the classical solution to the equations of motion of the linearized system. In the case of gravity such a split is not so easy to obtain. In this case one has to consider the eigenmodes of the Hessian of the action and the corresponding eigenvalues. The eigenvalues of the pseudo gauge modes will converge to zero in the perfect limit and will lead to a divergence of the measure.

Dropping the infinite volume of the gauge group, one arrives at the perfect propagator
\begin{eqnarray}\label{Gl:PerfectPropagatorQNeq0}
K(x_0,\tau_0,x_1,\tau_1)\;=\;\sqrt\frac{\omega}{2\pi\hbar\sinh(\omega T)}\exp\left(-\frac{1}{\hbar}S(x_0,\tau_0,x_1,\tau_1)\right)
\end{eqnarray}

\noindent where $S$ is given by (\ref{Gl:PerfectActionQNeq0}). Is is straightforward to check that the propagator (\ref{Gl:PerfectPropagatorQNeq0}) satisfies the constraint
\begin{eqnarray}
\hat C\,K(x_0,\tau_0,x_1,\tau_1)\;&=&\;\left(\hbar\frac{\partial}{\partial\tau_1}+\frac{\hbar^2}{2}\frac{\partial^2}{\partial x_1^2}+\frac{\omega^2}{2}x_1^2\right)K(x_0,\tau_0,x_1,\tau_1)\;=\;0
\end{eqnarray}

\noindent i.e. $K$ is a projector on the physical Hilbert space. Furthermore, writing explicitly the dependence of $K$ on the background solution $\bar q_i$ and $\bar t_i$, $K$ also satisfies
\begin{eqnarray}
&&K(x_0,\tau_0,x_2,\tau_2;\bar q_0,\bar t_0,\bar q_2,\bar t_2)\\[5pt]\nonumber
&&\qquad\qquad\;=\;\int dx_1\;K(x_0,\tau_0,x_1,\tau_1;\bar q_0,\bar t_0,\bar q_1,\bar t_1)K(x_1,\tau_1,x_2,\tau_2;\bar q_1,\bar t_1,\bar q_2,\bar t_2)
\end{eqnarray}

\noindent for every $\bar q_1$, $\bar t_1$ that satisfy
\begin{eqnarray}
\bar q_1\;=\;  \bar q_0 \frac{\sinh(\bar t_2-\bar t_1)}{\sinh(\bar t_2-\bar t_0)} + \bar q_2 \frac{\sinh(\bar t_1-\bar t_0)}{\sinh(\bar t_2-\bar t_0)}
\end{eqnarray}

\noindent i.e. that satisfy the equations of motion of the (Wick-rotated) continuum harmonic oscillator. In this sense the perfect propagator is independent of the actual discretization, i.e. for $N=2$ of which actual background solution one is perturbing around.
%
%
%
%
%
%
%
%

%

%
%

\section{The anharmonic oscillator}\label{Sec:AHO}

In section \ref{Sec:q=0} we dealt with the harmonic oscillator linearized around the background solution with $\overline{q}_n=0$. Here the linearization effectively lead to omitting the fluctuations $\tau_n$ in the time variables. Indeed, treating these explicitly in the path integral is quite complicated, as the time variables appear non-polynomially in the action.

To discuss a truly interacting theory we will now consider the quartic anharmonic oscillator, so that we need to expand the action to at least fourth order. We have to face the problem that the $\tau_n$ appear in the third and fourth order of the expansion. Nevertheless we will see that also in this case we can avoid the integration over the fluctuations $\tau_n$ by perturbing around the perfect propagator for the harmonic oscillator.

To start with, a discretization for the quartic anharmonic oscillator is given by the following action for one time step
\ba\label{b21}
S_{01}\;&=&\;\frac{1}{2}\frac{(q_{1}-q_0)^2}{(t_1-t_0)}\,+\,\frac{\omega^2}{4}(q_1^2+q_0^2) (t_1-t_0)\,+\,\frac{\lambda}{2 \cdot 4!}(q_0^4 + q_1^4)(t_1-t_0)  \q .
\ea
In a path integral approach we can treat such actions only perturbatively. For instance we can expand as before around the background solution $\overline{q}_n=0$ and $\overline{t}_n$ arbitrary. To second order in the perturbation variables $x_n, \tau_n$ we recover the harmonic oscillator, treated in section \ref{Sec:q=0}. In particular there the perturbations $\tau_n$ did not appear and we could just ignore the integration over these variables in the path integral.

To capture however the anharmonic term, we need to expand to higher order in the variables, which introduces a dependence on the $\tau_n$. Even worse, the lowest order terms in which the $\tau_n$ appear, are linear in the $\tau_n$,  which cannot be directly dealt with in the path integral.

Classically one can show \cite{hohn} that a perturbative treatment of actions with broken symmetries, which are however expanded around a background solution with exact symmetries, is in general inconsistent in higher-than-linear order. The problem is that the background gauge parameters (here the $\overline{t}_n$) are free to lowest order in perturbation theory but are fixed by (non-linear) consistency conditions to higher order. That is, if the $\overline{t}_n$ are not already chosen such that these consistency equations are satisfied, the ansatz $t_n=\overline{t}_n+ \epsilon \tau_n$ is not justified as the solutions for the $\tau_n$ will involve terms $\tau_n \sim \epsilon^{-1}$.

To avoid this issue one can improve the action order by order. That is  one would start with the quadratic order of the action and find the perfect action to this order. To this one adds the (naively discretized) third order term. For this third order action the problem above does not appear (to lowest non-linear order), i.e. the background gauge parameters $\overline{t}_n$ remain free. This third order action has to be improved again, after which one can add the fourth order term and so on.

This solves the consistency problem on the classical level. It also solves the problem in the path integral:  the action (and measure) being perfect up to a certain order will be also gauge invariant to this order. Hence one can either apply gauge fixing, or rewrite the action into gauge invariant variables, such that the path integral needs only be performed over these variables.

As explained in section \ref{Sec:q=0} in the case of the harmonic oscillator the perfect action for the quadratic approximation gives us already the full non-perturbative perfect action, if we just replace the background time parameters $\overline{t}_n$ with the full variables $t_n$. We therefore do not need to perform the perfection procedure to third order. To fourth order we have a contribution from the anharmonic potential term. So we will start from the action
\ba\label{b22}
S_{01}\;&=&\;\frac{\omega}{2}
\left(
\frac{\cosh( \omega(t_1-t_0))(q_0^2+q_1^2) - 2q_0q_1}{\sinh(\omega(t_1-t_0))}
\right)
\,+\,\frac{\lambda}{2 \cdot 4!}(q_0^4 + q_1^4)(t_1-t_0)  \q .
\ea
We will again consider the case with two time steps, so that we always need to integrate only over one variable pair $q_1,t_1$. That is we consider
\ba
S=S_{01}(q_0,t_0,q_1,t_1)+S_{12}(q_1,t_1,q_2,t_2)   \q .
\ea
Applying the expansion $q_n=0+x_n, t_n=\overline{t}_n+\tau_n$ we would actually encounter third and fourth order terms (from the expansion of the perfect part of the action) in which the $\tau_1$ appear. We know however that the perfect action is exactly gauge invariant, hence there exist a variable transformation $(q_1,t_1) \rightarrow (Q_1,T_1)$ such that this perfect part only depends on $Q_1$. Indeed with  $T_1=t_1$ and
\ba\label{b23}
Q_1:=\sqrt{\frac{\sinh(\omega(t_2-t_0))  }{ \sinh(\omega(t_1-t_0)) \sinh(\omega(t_2-t_1))    }}\,
\left(
q_1-\frac{ \sinh(\omega(t_2-t_1))}{\sinh(\omega(t_2-t_0) )} q_0-\frac{ \sinh(\omega(t_1-t_0))}{\sinh(\omega(t_2-t_0))} q_2
 \right) \q\q
\ea
 we find
\ba\label{b24}
S_{01}+S_{12}&=&\frac{\omega}{2}Q_1^2 +
\frac{\omega}{2}
\left(
\frac{\cosh( \omega(t_2-t_0))(q_0^2+q_2^2) - 2q_0q_2}{\sinh(\omega(t_2-t_0))}
\right) \nn\\
&&\,+ \frac{\lambda}{2 \cdot 4!}  \left( (q_0^4+ q_1^4(Q_1,q_0,q_2))(t_1-t_0) + (q_1^4(Q_1,q_0,q_2)+q_2^4)(t_2-t_1)\right)  \q .
\ea

In the last line $q_1$ has to be expressed as a linear combination of $Q_1,q_0$ and $q_2$ (with $t_n$ dependent coefficients) by inverting (\ref{b23}). Note that the second term in the action (\ref{b24}) is just Hamilton's principal function for the harmonic oscillator, coinciding with the perfect action for the time step $t_2-t_0$. We now expand the action (\ref{b24}) in $Q_1=0+X_1$, $q_n=0+x_n$ for $n=0,2$ and in $t_n=\overline{t}_n+\tau_n$. Indeed we see that $\tau_1$ does not appear in an expansion up to fourth order in the variables. We can therefore drop the $\tau_1$-integration in the path integral just as in the case of the harmonic oscillator. A similar argument can be applied if we consider the path integral involving any number $N$ of steps. Again one can find a transformation from the $q_1,\ldots,q_{n-1},t_1,\ldots,t_{n-1}$  to variables $Q_1,\ldots,Q_{N-1},t_1,\ldots, t_{N-1}$ such that the harmonic part of the action depends only on the $Q_k$. We can therefore ignore the fluctuation variables $\tau_k$ also in the iteration process (which computes the $N=2^M$ path integral iteratively) below.\footnote{Alternatively \cite{SebDipl} to introducing the variable $Q_1$ one can expand first the action in $q_n=0+x_n,t_n=\overline{t}_n+\tau_n$ for $n=0,1,2$ to fourth order and then define a coordinate transformation $x_1\rightarrow \tilde X_1$ such that the action to fourth order only depends on $\tilde X_1$ and not on $\tau_1$. This leads to the same results as the approach presented here.}\\[5pt]


This allows us to compute iteratively the perfect propagator for the anharmonic oscillator to first order in $\lambda$. As for the harmonic case we have to choose a parametrization for the propagator. To this end consider the first iteration step, i.e. integrating out $Q_1$ from (\ref{b24}), where again we assume $T=\overline{t}_2-\overline{t}_1=\overline{t}_1-\overline{t}_0$:
\ba\label{b25}
K^{(1)}(x_0,x_2,2T)&=&
\int dQ_1 \,( \eta(T))^2  \,\,  \exp\left( -\frac{\omega}{2\hbar} Q_1^2 -\frac{1}{\hbar} S_{harm}(x_0,x_2,2T) \right) \times \nn\\
&&\q\q\q\q\q\q\q
\left( 1- \frac{\lambda}{\hbar}\Big(I^{(0)}_{01}(T)+I^{(0)}_{12}(T)\Big)+O(\lambda^2) \right)  \q\q\q  \nn\\
&=:& \eta(2T) \exp\left( -\frac{1}{\hbar} S_{harm}(x_0,x_2,2T) \right)\left(1-\frac{\lambda}{\hbar} I^{(1)}_{02}(2T)+O(\lambda^2) \right)
\ea
The action $S_{harm}$ is the perfect action for the harmonic oscillator appearing in (\ref{b24}). It does not depend on $Q_1$, hence this exponential factor can be pulled out of the integral.

We have chosen a measure factor
\ba\label{b26}
\eta(T)=\sqrt{\frac{\omega}{2\pi\hbar \sinh (T \omega)}} \left(\frac{\sinh (T\omega)}{2\cosh (T \omega)}\right)^{1/4}
\ea
which corresponds to the perfect measure for the harmonic oscillator for the $Q_1$ variable. The second factor is (the square root of) the Jacobian of the transformation from $q_1$ to $Q_1$. To zeroth order in $\lambda$ this measure factor remains indeed invariant under iteration. (First order corrections in $\lambda$ to the measure can be absorbed into the interaction term $I_{02}$.)

The interaction terms $I^{(0)}_{01}$, $I^{(0)}_{12}$ are polynomials of up to fourth order in $X_1$. The coefficients of the $X_1$ in these polynomials depend on $x_0,x_2$ in such a way that all terms are fourth order if we add up the powers in $x_0$, $X_1$, and $x_2$. Performing the integration\footnote{$\int dX  \, X^{2k} \exp(- a x^2 )=a^{-(k+\frac{1}{2})}\Gamma(k+\frac{1}{2})$.} in (\ref{b25}) one will find that $I^{(1)}_{02}$ will contain all even powers in $x_0$, $x_2$ up to fourth order including a constant term. We therefore adopt the following parametrization for the interaction term
\ba\label{b27}
I_{01}&=& \alpha_0(T)\,(x_0^4+x_1^4) +\alpha_1(T)\,(x_0^3x_1+x_0x_1^3) +\alpha_2(T) x_0^2x_1^2 \,\,+\nn\\
&&\beta_0(T) \,(x_0^2+x_1^2) +\beta_1(T) \, x_0x_1  \,+\, \gamma(T) \q .
\ea
From the iteration
\ba\label{b28}
K^{(n+1)}(x_0,x_2,2T)\;=\;\int dX_1\,\,K^{(n)}(x_0,x_1(X_1),T)\,\,K^{(n)}(x_1(X_1),x_2,T) + O(\lambda^2)
\ea
with
\ba\label{b29}
K^{(n)}(x_0,x_1,T)=\eta(T) \exp\left( -\frac{1}{\hbar} S_{harm}(x_0,x_1,T) \right)\left(1-\frac{\lambda}{\hbar} I^{(n)}_{01}(T)+ O(\lambda^2)\right)
\ea
\noindent we obtain recursion relations for the coefficients $\alpha$, $\beta$, and $\gamma$. This is a straightforward exercise involving only the transformation from $x_1$ to $X_1$ given by (\ref{b23}), just replacing $q_1$, $Q_1$ by $x_1$, $X_1$ and performing the integration over $X_1$.

By expanding in powers of $\hbar$ these recursion relations can be divided into a classical part and quantum corrections. The classical part of the recursion relations coincides with the recursion relations one would obtain by extremizing
\ba\label{30}
S_{harm}(x_0,x_2,2T)+\lambda I^{(n+1)}(x_0,x_2,2T) &=& S_{harm}(x_0,x_1,T) +S_{harm}(x_1,x_2,T) + \nn\\&&\q\q \lambda I^{(n)}(x_0,x_1,T)+\lambda I^{(n)}(x_1,x_2,T) +O(\lambda^2)\q\q\q
\ea
with respect to $x_1$. A solution to the corresponding fixed point equations is given by Hamilton's principal function (for the corresponding continuum action), which can be readily obtained to first order in $\lambda$, see appendix \ref{app1}.

We will see that the classical part coincides with the homogeneous part of the recursion relations. As will be explained in the appendix \ref{App:BetaRecursion}, the inhomogeneous recursion relations can be easily brought into a standard form once the fixed point solutions to the homogeneous part is known.  Hence it helps very much to have the classical system solved, in order to obtain the full quantum mechanical solution.

The recursion relations for the $\alpha_i$-coefficients are a closed system as these correspond to the classical problem with just an $x^4$ order interaction term:
\begin{eqnarray}\label{b31}
\alpha_0^{(n+1)}(2T)\;&=&\frac{1}{8\cosh^4(T\omega)}\bigg( (1+ 8 \cosh^4(T\omega))\, \alpha_0^{(n)}(T) + (4\cosh^3(T\omega)+\cosh(T\omega))\,\alpha_1^{(n)}(T)  + \nn\\&&\q\q\q\q\q\q\q 2\cosh^2(T\omega)\,\alpha_2^{(n)}\bigg)\nn\\
  \alpha_1^{(n+1)}(2T)\;&=&\frac{1}{2\cosh^4(T\omega)}\bigg(    \alpha_0^{(n)}(T) + (\cosh^3(T\omega)+ \cosh(T\omega)  )\,\alpha_1^{(n)}(T) + \cosh^2(T\omega)\,\alpha^{(n)}_2 \bigg)
  \nn\\
\alpha_2^{(n+1)}(2T)\;&=&\;\frac{1}{4 \cosh^4(T\omega)} \left(3 \,\alpha_0^{(n)}(T) + 3 \cosh(T \omega) \,\alpha_1^{(n)}(T) +
   2 \cosh^2(T \omega) \,\alpha_2^{(n)}(T)\right) \q .
\end{eqnarray}

One fixed point of these equations is provided by the perfect action for the anharmonic oscillator (to linear order in $\lambda$), determined in  the appendix \ref{app1}:
\begin{eqnarray}\label{b31a}
\alpha_0^{*}(T)\;&=&\;\frac{\tilde \lambda}{768 \omega \sinh^4(T \omega)}\Big(12 T \omega - 8 \sinh(2 T \omega) + \sinh(4 T \omega)\Big)\nn\\
\alpha_1^{*}(T)\;&=&\;\frac{\tilde \lambda}{192 \omega \sinh^4(T \omega)}\Big(-12 T \omega \cosh(T \omega)+9 \sinh(T \omega)+ \sinh(3 T \omega)\Big)\nn\\
\alpha_2^{*}(T)\;&=&\;\frac{\tilde \lambda}{64 \omega \sinh^4(T \omega)}\Big(2 T \omega (2 + \cosh(2 T \omega)) - 3 \sinh(2 T \omega)\Big)
\end{eqnarray}
The fixed point equations are invariant under a rescaling $\alpha_i \rightarrow  \tilde \lambda \alpha_i$, and hence we have $\tilde \lambda$ as a free parameter for the solutions. This is easy to understand as the freedom to rescale the interaction term by an arbitrary constant, redefining the coupling constant $\lambda$. As for the ambiguities $g$, $\tilde \omega$ appearing in the fixed points for the harmonic oscillator, the final coupling constant is determined by the initial conditions for the iteration procedure.

The solution (\ref{b31a}) is actually not the most general one. We will explain in the appendix \ref{unique} that the most general solution has two further free parameters, which determine the couplings to  terms $\dot x^4$ and $\dot x^2 x^2$ in the corresponding continuum Lagrangian. In the following we will however set these couplings to zero, that is consider the standard quartic anharmonic oscillator with a perturbation term $x^4$.


Next we consider the recursion relations for the $\beta_i$, which are given by
\begin{eqnarray}\label{b32} \nn
\beta_0^{(n+1)}(2T)\;&=&\;\beta_0^{(n)}\left(T\right)\left(1+\frac{1}{2
\cosh^2(T\omega)}\right)\;+\;\beta_1^{(n)}\left(T\right)\frac{1}{2
\cosh(T \omega)}\\\nn
&&\;+\;
\frac{\hbar}{\omega} \Big( \frac{3 \tanh(T \omega)}{2 \cosh^2(T \omega)} \alpha_0^{(n)}(T) + \frac{3 \tanh(T \omega)}{4 \cosh(T \omega)} \alpha_1^{(n)}(T) + \frac{\tanh(T\omega)}{2} \alpha_2^{(n)}(T) \Big)
\nn\\
\beta_1^{(n+1)}(2T)\;&=&\;\beta_0^{(n)}\left(T\right)\frac{1}{\cosh^2(T\omega)}\;+\;\beta_1^{(n)}\left(T\right)\frac{1}{
\cosh(T\omega)} \nn \\
&&\;+\;
\frac{\hbar}{\omega}\Big(\frac{3 \tanh(T \omega)}{\cosh^2(T \omega)} \alpha_0^{(n)}(T)+ \frac{3 \tanh(T \omega)}{2 \cosh(T\omega)} \alpha_1^{(n)}(T)\Big)  \q .
%
\end{eqnarray}

\noindent The equations (\ref{b32}) have a homogenous part which is independent of $\hbar$, and an inhomogeneity proportional to $\hbar$ depending on the $\alpha_i$, i.e.~the fourth order potential.  To any solution of the full recursion relations one can add an arbitrary multiple of the solutions to the homogeneous part of the recursion relations. Hence to find all solutions one has to determine also the solutions to the homogeneous part. As this part  is of the order of $\hbar^0$ it is again equivalent to the classical iteration relations one would obtain if one considers an interaction term quadratic in $x$ added to the perfect action for the harmonic oscillator. A solution for the homogeneous fixed point equations is hence again given by the perfect action for the harmonic oscillator: Take (\ref{fp1}) and replace there $\tilde \omega,g \rightarrow \omega +\lambda \tilde\nu, 1+\lambda \tilde\mu$ and expand to first order in $\lambda$, resulting in
\begin{eqnarray}\label{b33}
\beta_0^{h}(T)\;&=&\;\mu \coth(T \omega) - \frac{\nu T \omega}{\sinh^2(T\omega)} \nn\\
\beta_1^{h}(T)\;&=&\;\nu \frac{2 T \omega \cosh(T \omega)}{\sinh^2(T \omega)} - \mu \frac{2}{\sinh(T \omega)} \q .
\end{eqnarray}
(Here we reparametrized $\mu=\tfrac{1}{2}(\tilde \nu-\omega \tilde \mu)$ and $\nu=\tfrac{1}{2} \tilde \nu$.)
Again we find a two--parameter ambiguity corresponding to the ambiguities $g$ and $\tilde \omega$ that we found for the harmonic oscillator.

For the inhomogeneous equations we can assume that the $\alpha_i$ are given by their fixed point values (\ref{b31a}). There is a general strategy with which one can attempt to find the fixed points for the inhomogeneous equations, which requires the knowledge of the homogeneous equations.  We will demonstrate this in the appendix \ref{App:BetaRecursion} as an example. The general solution is given by
\begin{eqnarray}\label{b33a}
\beta_0^{*}(T)\;&=&\;\mu \coth(T \omega)\,-\,\frac{\nu T \omega}{\sinh^2(T\omega)}\nonumber\\\nonumber
&&\;+\;\frac{\tilde \lambda\hbar}{32 \omega^2 \sinh^2(T \omega)}\Big(2 + \cosh^2(T \omega) - 3 T \omega \coth(T \omega)\Big)\\
\beta_1^{*}(T)\;&=&\;\nu\frac{2T \omega \cosh(T \omega)}{\sinh^2(T \omega)}\,-\,\mu
\frac{2}{\sinh(T \omega)}\nn\\
&&\;+\;\frac{\tilde \lambda\hbar}{32 \omega^2 \sinh^3(T \omega)}\Big(4 T \omega + 2 T \omega \cosh(2 T \omega) - 3 \sinh(2 T\omega)\Big)  \q .
\end{eqnarray}


Finally the recursion relations for $\gamma$ are given by
\begin{eqnarray}\label{Gam}
\gamma^{(n+1)}(2T)\;&=&\;2\gamma^{(n)}(T)\,+\frac{\hbar \tanh(T \omega)}{\omega} \beta_0^{(n)}(T)+ \frac{3 \hbar^2 \tanh^2(T \omega)}{2 \omega^2} \alpha_0^{(n)}(T) \q .
\end{eqnarray}
Again we have a homogeneous part corresponding to a classical iteration procedure. This corresponds to having a constant potential added to the perfect action for the harmonic oscillator. It is easy to see that under the classical recursion relations such a constant term is just multiplied by $2$ corresponding to changing the time interval from $T$ to $2T$. Hence a (family of) fixed points for the homoegenous part of the relations is just given by $\gamma^*(T)_{hom}=\xi T$.

The fixed points for $\gamma^*(T)$ for the full recursion relations can be readily found with methods shown in appendix \ref{App:BetaRecursion}. These are given by
\begin{eqnarray}\label{Gams}
\gamma^{*}(T)\;&=&\;\xi T\,-\,\frac{\hbar}{\omega}\Big(\mu-\nu T\omega \coth(T \omega)\Big)- \frac{\tilde \lambda\hbar^2}{64 \omega^3}\Big(3 \coth(T \omega) - T \omega (2 + \frac{3}{\sinh^2(T \omega)})\Big) \q .\q\q
\end{eqnarray}


To summarize, the fixed point propagator is
\begin{align}\label{b41}
&K^{(\mu,\nu,\xi,\tilde{\lambda})}(x_0,x_1,T)=\nonumber \\
 =& \sqrt{\frac{\omega}{2 \pi \hbar \sinh(T \omega)}} \exp\Big(-\frac{1}{\hbar} S_{harm}(x_0,x_2,T)\Big) \times
 \Bigg(1 - \frac{{\lambda}}{\hbar} \times \nn\\
&  \Bigg[
 \frac{\tilde \lambda}{768 \omega \sinh^4(T \omega)} \Big(12 T \omega - 8 \sinh(2T\omega) + \sinh(4 T \omega)\Big) (x_0^4 + x_1^4)+ \nonumber \\
& \frac{\tilde \lambda}{192 \omega \sinh^4(T \omega)}\Big(-12 T \omega \cosh(T \omega) + 9 \sinh(T\omega) +\sinh(3 T \omega)\Big) (x_0^3 x_1 + x_0 x_1^3) + \nonumber \\
& \frac{\tilde \lambda}{64 \omega \sinh^4(T \omega)}\Big(2 T \omega (2 + \cosh(2 T \omega)) - 3\sinh(2 T \omega)\Big) x_0^2 x_1^2 + \nonumber \\
& \Big( \mu \coth(T \omega) -  \frac{\nu T \omega}{\sinh^2(T \omega)} + \tilde \lambda \hbar \frac{\Big(2 + \cosh^2(T \omega) -  3 T\omega \coth(T \omega)\Big)}{32 \omega^2 \sinh^2(T \omega)}  \Big)(x_0^2 + x_1^2) + \nonumber \\
& \Big(\nu \frac{2 T \omega \cosh(T \omega)}{\sinh^2(T \omega)} - \mu \frac{2}{ \sinh(T \omega)}+
\tilde \lambda\hbar  \frac{\Big(4 T\omega + 2 T \omega \cosh(2 T \omega) - 3 \sinh(2 T\omega)\Big)}{32 \omega^2 \sinh^3(T \omega)}\Big) x_0 x_1 + \nonumber \\
& \xi T  - \frac{\hbar}{\omega} \Big( \mu - \nu T \omega \coth(T \omega)\Big) - \frac{\tilde \lambda\hbar^2}{64 \omega^3}\Big(3 \coth(T \omega) -  T \omega (2 + \frac{3}{\sinh^2(T \omega)})\Big) \Bigg] + O(\lambda^2) \Bigg) \q .
 \end{align}

For the initial data used in (\ref{b25}) the propagator takes on a simplified form, as in this case $\tilde \lambda=1$ and $\mu=\nu=\xi=0$. Nevertheless the result is quite complicated: even to guess the correct reparametrization invariant measure (i.e. all terms proportional to $\hbar$ and $\hbar^2$ in square brackets) in case the perfect classical action is given, without actually solving the dynamics, seems to be quite impossible in general situations.

From the infinitely many parameters $\alpha_i(T)$, $\beta_i(T)$, and $\gamma(T)$ in our initial parametrization we are left with the couplings $\tilde \lambda$, $\mu$, $\nu$, and $\xi$ (and a further two couplings corresponding to adding terms $\dot x^4$ and $\dot x^2 x^2$ to the continuum Lagrangian, see appendix \ref{unique}).  These parameters characterize the continuum Lagrangian (and Hamiltonian, see below), i.e.~all discretization ambiguities are resolved by requiring reparametrization invariance for the discretized path integral.\\[5pt]



Again, although for the recursion equations (\ref{b28}) we have set $t_1=\frac{t_0+t_2}{2}$, the fixed points of the recursion equations satisfy a stronger condition, namely
\begin{eqnarray}\label{Gl:RecursionAHOStronger}
K(x_0,x_2,T_1+T_2)\;=\;\int dx_1\,K(x_0,x_1,T_1)K(x_1,x_2,T_2)\;+\;O(\lambda^2)  \q
\end{eqnarray}
for arbitrary positive $T_1$, $T_2$. In other words, the perfect propagator $K(x_0,x_1,T)$ leads to a path integral which for $N>1$ has become independent -- to first order in $\lambda$ -- on the actual placement of the intermediate discretization points, which correspond to the discretized choice of $t(s)$ in (\ref{Gl:ActionParameterizedContinuous}). It therefore mimics exactly the gauge symmetry of the continuum theory. Also (\ref{Gl:RecursionAHOStronger}) shows that the model defined by the propagator (\ref{b41}) is discretization independent (up to terms of order $\lambda^2$), i.e.~the discrete path integral defined by the amplitude (\ref{b41})  does not depend on the number of discretization points.

The invariance under a gauge symmetry leads to a constraint, which is satisfied by the propagator (\ref{b41})
\begin{eqnarray}
\hat{C}^{(\mu,\nu,\xi,\tilde{\lambda})} := \hbar\frac{\partial}{\partial T} + (1+2(\nu - \mu))\frac{\hbar^2}{2}\frac{\partial^2}{\partial x_1^2} + \frac{x_1^2}{2}(\omega^2 +2  (\nu + \mu)) + \frac{\tilde{\lambda}\lambda }{4!} x_1^4 +  \xi
\end{eqnarray}

\noindent such that
\begin{eqnarray}\label{Gl:ProjectorOnPhysicalHilbertSpaceAHO}
\hat C^{(\mu,\nu,\xi,\tilde{\lambda})}K(x_0,x_1,T)\;=\;O(\lambda^2)     \q .
\end{eqnarray}
This constraint equation characterizes the fixed points, it would not be satisfied by the $K^{(n)}$ for finite $n$.

\section{Summary and Discussion}\label{Sec:Summary}

In order to obtain a well defined path integral one often utilizes discretizations. This may however lead to a breaking of (gauge) symmetries, which in the case of general relativity are central for the dynamics of the system. We addressed this problem here in the context of reparametrization invariant systems. Despite being extremely simple compared to gravity, these systems share the property that the gauge symmetry determines the dynamics of the system \cite{kuchar}.

Indeed we could show in section \ref{sec2} that requiring the implementation of reparametrization invariance for the discretized path integral -- where it assumes the form of vertex translations (\ref{b04}) -- uniquely fixes the discrete propagator.  Namely the propagator has to coincide with the quantum mechanical continuum propagator of the system under consideration. Furthermore such a discretized path integral is automatically discretization independent, i.e.~the propagator does not depend on the number of subdivisions . This is just the convolution property (\ref{b07}) of the continuum quantum mechanical propagator. The number of subdivisions can be taken to zero, in which case one obtains the discrete propagator itself, even for large time steps. This (fixed point) discrete propagator is furthermore characterized by satisfying the constraint equations.

We conjecture that similar properties will also hold for discrete gravity: a full implementation of diffeomorphism invariance in the form of a symmetry under vertex translations \cite{Bahr:2010cq} should lead to discretization independence. Similar to the example of reparametrization invariance, the discretization can then be taken to be a very coarse one, hence one would expect that many discretization ambiguities will be fixed (as free parameters in the discrete propagator have an effect on macroscopic scales). It would be interesting to have a proof for more general cases, in analogy to the arguments presented here. One important difference to the one-dimensional case is that a perfect action or perfect discretization for higher dimensional theories will include non-local couplings \cite{Hasenfratz:1997ft, Bietenholz:1999kr,Bahr:2010cq}.

The implementation of vertex translation symmetry into the discretized path integral will also ensure that this path integral satisfies constraints \cite{hartle}, which can be taken as a characterization of the fixed point. Again, an important difference to the one-dimensional case is, that such constraints will involve non--local couplings, and are therefore not explicitly known. These constraints would however be free of anomalies.

Coming back to reparametrization invariant systems, we can take the requirement of discretization independence in the form of the convolution property (\ref{b07}) as a starting point to define an iterative procedure.  This iterative procedure improves the discrete propagator, so that in the limit it satisfies the convolution property (\ref{b07}) and defines a discretization in which reparametrization invariance is realized in the form of vertex translations. Some points we wish to emphasize, are the following:
\begin{itemize}
\item Quite naturally, one is led to consider not one specific discretization, but an entire class of discrete models. The iterative procedure defines a renormalization group flow within this class. This allows a discussion of all possible fixed points. Indeed here, we found not only the solutions to the (an--) harmonic oscillator, but to the most general Lagrangian involving even powers in $q$ and $\dot q$ up to order $4$. Hence to study the relevance of discretization ambiguities in gravity models \cite{Perez:2005fn}, it might be essential to study the behavior of these models under coarse graining. Note also that the relevant parameters -- which fix the corresponding continuum Lagrangian -- are all determined by the lowest orders in an expansion of the discrete propagator in the time variable, that is by the behavior of the propagator for short times. The higher order coefficients in this expansion can indeed be understood as proper discretization ambiguities, which are fixed by requiring reparametrization invariance of the discrete propagator. This might be an interesting point for spin foam models, as there one often rather concentrates on the limit of having large building blocks  (with an amplitude that would correspond to what we termed naive discretization, as it has not been subject to any coarse graining procedure).

\item Even for the simple case of the anharmonic oscillator, the reparametrization invariant discrete path integral (\ref{b41})  is quite complicated, and it would probably be impossible to guess it without actually solving the dynamics. Similarly, to find an anomaly-free measure for spin foam models \cite{Bojowald:2009im, Bianchi:2010fj, Bahr:2010my}, it seems to be unavoidable to address the dynamics of the system, in particular the behavior under coarse graining. Even to just ensure anomaly-freeness in the continuum limit (which is the only thing one might realistically expect), a study of the coarse graining properties might be valuable. As we have seen, the choice of initial data for the iterative procedure of the measure factor had to be done carefully in order to obtain a convergent result. The behavior of the amplitudes under (dynamically) trivial subdivisions \cite{Bojowald:2009im, Bahr:2010my} are nevertheless interesting as first steps in this direction. These might actually be important in order to show that vertex translation symmetry implies discretization independence for gravity, in analogy to the one-dimensional case.

\item Path integrals with (properly implemented) gauge symmetries act
as projectors onto the space of physical states \cite{hartle}. Indeed
spin foams are often mentioned as a tool to obtain physical states, or
to define the physical inner product, for loop quantum gravity. As we
have seen in the toy example of reparametrization invariant systems,
to obtain a propagator satisfying the quantum constraints, it was
necessary to take the fixed point propagator, that is the perfect or
continuum limit of the (naively) discretized path integral. On the
other hand the constraints can be expressed as conditions on the fixed
point coefficients $\alpha_i^*(T),\ldots,\gamma^*(T)$. For future work
it would be interesting to explore the relation between these
conditions encoding reparametrization invariance, and the fixed point
conditions encoding discretization independence.

\item Finally, the methods applied in this work -- basically a version of Wilsonian Renormalization Group flow -- might be actually useful in order to find solutions to quantum mechanical path integrals. Also it would be interesting to apply these techniques to
reparametrization invariant systems arising in mini-superspace
reductions of gravity, for instance for loop quantum cosmology
\cite{lqc}. The path integral for the anharmonic oscillator is usually used to derive the corrections to the energy levels of the harmonic oscillator \cite{MacKenzie:1999pu}, for which one just needs to obtain the $T\rightarrow \infty$ behavior of the propagator. Here we derived the full propagator for arbitrary $T$, as this is needed to define a perfect discretization of the path integral, and with this an anomaly-free (with respect to vertex translation symmetry) path integral measure. We also want to point out the work \cite{Bietenholz:1997rt}, where the authors also discussed a perfect path integral for the anharmonic oscillator. The difference to the work presented here is that  \cite{Bietenholz:1997rt} uses a different coarse graining procedure, namely averaging instead of decimation as applied here. We employed decimation as this seems to be the only method to obtain reparametrization invariance.

\end{itemize}

\section*{Acknowledgements}

The authors would like to thank Wolfgang Bietenholz and Carlo Rovelli for discussions about perfect actions. BB would like to thank Professor Horgan for discussions about fixed point actions.
\appendix

\section{Hamilton's principal function for the anharmonic oscillator}\label{app1}

Here we give Hamilton's principle function for the anharmonic oscillator to first order in $\lambda$, as this will provide a fixed point for the recursion relations for $\alpha_i$ (\ref{b31}).

In \cite{bahrdittrich1} it was shown that for a $1D$ system the perfect action coincides with Hamilton's principal function for the given boundary values, i.e.
\begin{eqnarray}\label{Gl:HamiltonJacobiAHO}
S_{perf}(q_0,t_0,q_1,t_1)\;=\;\int ds\,\left( \frac{1}{2}\frac{(q')^2}{ t'}+\frac{\omega^2}{2} q^2 t'+\frac{\lambda}{4!} q^4 t'\right)
\end{eqnarray}

\noindent where $q(s)$ and $t(s)$ are solutions to the continuum equations of motion with boundary values $q_0$, $t_0$, $q_1$, and $t_1$.  To find Hamilton's principal function to first order in $\lambda$ we would need to find the solutions at most to first order in $\lambda$, and then perform the integral in (\ref{Gl:HamiltonJacobiAHO}),  i.e. we expand
\begin{eqnarray}\label{Gl:AHOSolution}
q(s)\;=\;\bar q(s)\,+\,\lambda x(s)\,+\,O(\lambda^2)\\[5pt]
t(s)\;=\;\bar t(s)\,+\,\lambda\tau(s)\,+\,O(\lambda^2)
\end{eqnarray}

\noindent where $\bar q$, $\bar t$ are solutions to the harmonic oscillator. However to first order in $\lambda$ we do not need the explicit form of the solutions $x(s)$, $\tau(s)$: These would only appear in the harmonic oscillator part of the action (\ref{Gl:HamiltonJacobiAHO}). This contribution vanishes however due to the (harmonic oscillator) equations of motion for the background solution $\bar q(s)$, $\bar t(s)$. Hence we just need to evaluate the integral (\ref{Gl:HamiltonJacobiAHO}) on the harmonic oscillator solution. One finds
\begin{eqnarray}\label{Gl:PerfectActionAHO}
S_{perf}(q_0,t_0,q_1,t_1)
\;&=&\;\frac{\omega}{2}\frac{\cosh(T \omega) (q_0^2+q_1^2)-2q_0q_1}{\sinh(T \omega)} \; +  \nn\\
&&\frac{\lambda }{768 \omega \sinh^4(T \omega)}\bigg[\bigg(12T \omega - 8\sinh(2T \omega) + \sinh(4T\omega) \bigg)(q_0^4+q_1^4)\;+\nn\\
&&\bigg(-48T\omega \cosh(T \omega) + 36 \sinh(T \omega) + 4\sinh(3T \omega) \bigg)(q_0q_1^3+q_0^3q_1)\;+\nn\\
&&\left.\bigg(24T\omega (2 + \cosh(2T\omega)) - 36\sinh(2T \omega)\bigg)q_0^2q_1^2\right]  \q .
\end{eqnarray}
Similarly one can obtain Hamilton's principal function to first order in $\lambda$ for the harmonic oscillator with perturbation terms  $\lambda \dot q^4$ and $\dot q^2 q^2$. As explained in the next section, these terms arise in the most general fixed point solution to the recursion relations (\ref{b31}).


\section{On the uniqueness of the fixed point solutions} \label{unique}

Here we will discuss the uniqueness of the solutions to the fixed point equations (\ref{Gl:HORec1}), (\ref{Gl:HORec2}), (\ref{Gl:HORec3}), (\ref{b31}) in the main text.
In all cases we assume that the the solutions can be represented by a power series $\sum_{n=n_0}^\infty c_n T^n$, which starts with some finite lowest power $T^{n_0}$, that can also be negative.

We start our considerations with the relations  (\ref{Gl:HORec1}) - (\ref{Gl:HORec3}) for the discretized action of the harmonic oscillator, which we rewrite into
\ba\label{un1}
0&=&     \alpha_1^*(2T)\;\alpha_1^*(T) -  \alpha_1^*(T)^2 + \tfrac{1}{8}\alpha_2^{*}(T)^2\\[5pt]
0&=& \alpha_2^*(2T)\;\alpha_1^*(T)  + \tfrac{1}{4}\alpha_2^*(T)^2   \q .
\ea

Making the ansatz
\ba\label{un2}
\alpha_i(T)&=& \sum_{n=n_0}^\infty \alpha_{i,n} T^n
\ea
one will find the following equation arising from the coefficients to the lowest power $T^{-2n_0}$  in the equations (\ref{un1})
\ba\label{un3}
0&=&\; (2^{n_0}-1)(\alpha_{1,n_0})^2 +\tfrac{1}{8}(\alpha_{2,n_0})^2  \nn\\
0&=&\; 2^{n_0}\,\alpha_{1,n_0}\,\alpha_{2,n_0} +\tfrac{1}{4} (\alpha_{2,n_0})^2   \q .
\ea
It is easy to see that this equation can be only solved for $n_0=-1$, in which case we obtain
\be\label{un3a}
\alpha_{2,-1}=\kappa_1 \; ,\q\q \alpha_{1,-1}=-\tfrac{1}{2} \kappa_1  \q
\ee
where $\kappa_1$ is a free parameter. After having fixed the lowest order, one can convince oneself that if one iteratively solves the higher order equations for  the coefficients of $T^{-2+k+1}$, then these equations are linear (inhomogeneous) equations for the coefficients $\alpha_{i,k}$, i.e. are of the form
\ba\label{un4}
\sum_j A_{ij} \alpha_{j,k} &=& h_i  \q .
\ea
Here the matrix $A$ is given by
\be\label{un5}
A = \left( \begin{array}{cc}  \q 2^{k-1}-\tfrac{3}{4}  \q& \q -\tfrac{1}{4}\q  \\ -\tfrac{1}{2} & 2^{k-1}-\tfrac{1}{2} \\ \end{array} \right)
\ee
and $h_i$ represents the inhomogeneous terms. The matrix has only vanishing determinant for $k=1$ (and $k=-1$, which we already discussed). Indeed, the equations for $\alpha_{i,1}$ add a further free parameter $\kappa_2$, as one will find the solutions
\ba\label{un6}
\alpha_{2,1}=\kappa_2 \; ,\q\q \alpha_{1,1}=-\kappa_2 \q .
\ea

Since the matrix $A$ has non-vanishing determinant for all other $k$'s one will have unique solutions for all the other coefficients depending on the two free parameters $\kappa_1,\kappa_2$. The solution obtained in this way agrees with the one (\ref{b2}) presented in the main text, with an appropriate choice of the two free parameters $g$, $\tilde \omega$ there.

One will find the same matrix $A$ appearing in the recursion relations (\ref{b32}) for the $\beta$-coefficients in the anharmonic oscillator case. This is not surprising, as these arise by adding a perturbation to the harmonic oscillator quadratic in the variables $x_0$, $x_1$. Since the determinant of the matrix $A$ is only vanishing in two cases $k=-1,1$ one will again find at most solutions with two free parameters, which is indeed the case for the solutions (\ref{b33a}). \\[5pt]

We will now discuss the solutions to the fixed point equations (\ref{Gl:HORec3})
\ba\label{et1}
\eta^*(2T) \; &=& \;\sqrt{\frac{\pi\hbar \sinh(\tilde \omega T)}{2 \cosh(\tilde \omega T)}} \eta^*(T)^2
\;\;\;\;=\;\;\, \left[ T^{1/2}   \sum_{k=0}^\infty  c_{2k}T^{2k}\right]  \, \eta^*(T)^2 \q .
\ea
Here we used in the fixed point solution (\ref{Gl:HORec3}) and expanded the prefactor appearing in the first equation into a power series.        As we have to choose for our initial measure $\eta^{(0)}$ a functional dependence $\eta^{(0)} \sim T^{-1/2}$ ,we start with the assumption that $\eta^*$ is of the form
\ba\label{et2}
\eta^*(T)&=& T^{-1/2} \sum_{n=n_0}^{\infty} \eta_n T^n \q
\ea
with some finite, not necessarily positive number $n_0$. Using this form in the fixed point conditions (\ref{et1}), it is easy to see that we must have $n_0=0$ and that the first coefficient $\eta_0$ is given by
\ba\label{et3}
\eta_0=2^{-1/2}c_0^{-1} \q .
\ea
Using (\ref{et3}), one will find that the coefficient equation for the power of $T^{1-1/2}$ in (\ref{et1}) leaves $\eta_1$ as a free parameter. Indeed for the coefficients $\eta_k$ with $k>0$, the equations are of the form
\ba\label{et4}
(2^k-2)\eta_k &=& f_k(\eta_l,\,l<k)  \q ,
\ea
with $f_1=0$. Hence $\eta_1$ remains a free parameter, determining all other coefficients uniquely.
The additional parameter corresponds to adding a constant potential term  to the Lagrangian for the harmonic oscillator. The full solution is given in (\ref{etas}).
\\[5pt]
%
%

Now we will turn to the recursion relations (\ref{b31})
\ba\label{un7}
\alpha_0^{*}(2T)\;&=&\frac{1}{8\cosh^4(T\omega)}\bigg( (1+ 8 \cosh^4(T\omega))\, \alpha_0^{*}(T) + (4\cosh^3(T\omega)+\cosh(T\omega))\,\alpha_1^{*}(T)  + \nn\\&&\q\q\q\q\q\q\q 2\cosh^2(T\omega)\,\alpha_2^{*}\bigg)\nn\\
  \alpha_1^{*}(2T)\;&=&\frac{1}{2\cosh^4(T\omega)}\bigg(    \alpha_0^{*}(T) + (\cosh^3(T\omega)+ \cosh(T\omega)  )\,\alpha_1^{*}(T) + \cosh^2(T\omega)\,\alpha^{*}_2 \bigg)
  \nn\\
\alpha_2^{*}(2T)\;&=&\;\frac{1}{4 \cosh^4(T\omega)} \left(3 \,\alpha_0^{*}(T) + 3 \cosh(T \omega) \,\alpha_1^{*}(T) +
   2 \cosh^2(T \omega) \,\alpha_2^{*}(T)\right)
\ea
 for the $x^4$ terms in the discretized action of the anharmonic oscillator.  This case is even easier to treat than the relations (\ref{un1}) as we have a system linear in the variables.  We assume again an ansatz
\ba\label{un8}
\alpha_i(T)&=& \sum_{n=n_0}^\infty \alpha_{i,n} T^n
\ea
with a lowest order $T^{n_0}$. Since the $\cosh(T\omega)^{-l}$ functions appearing in (\ref{un7}) can be expanded in a Taylor series (i.e. there are no negative powers of $T$ appearing), the equation for the lowest order coefficients $\alpha_{i,n_0}$ will be
\ba\label{un9}
\sum_{j}   A_{ij}\alpha_{j,n_0} &=&0
\ea
where
\ba\label{un10}
A= \left( \begin{array}{ccc}  \q   \tfrac{9}{8}-2^k \q&\q \q \tfrac{5}{8}\q\q    &  \q \q  \tfrac{1}{4}\q\q \\ \tfrac{1}{2} & 1-2^k & \tfrac{1}{2}\\
\tfrac{3}{4}  &\tfrac{3}{4} & \tfrac{1}{2}-2^k
 \end{array} \right)  \q .
\ea
The determinant of this matrix is only vanishing for $k=-3$, $-1$ and $k=1$ and for these cases the matrix has rank $2$. Hence we can expect three linearly independent solutions. For the higher order coefficients one has the same equation as in (\ref{un9}), just that inhomogeneous terms (arising from the lower order coefficients) have to be added to the right hand side. As the determinant of $A$ is non-vanishing except for the three cases mentioned above, we do not have any further free parameter than the three parameters from the three linearly independent solutions.

The solutions with lowest order $n_0=1$ is the one (\ref{b31a}) displayed in the main text. The most general solution is
\ba\label{un10a}
\alpha_0(T)\;&=&\;\frac{1}{768 \omega \sinh^4(T \omega)}\Big[4 (3\tilde{\lambda}_{-3} - \tilde{\lambda}_{-1} + 3 \tilde{\lambda}_1) T \omega + 8 (\tilde{\lambda}_{-3} - \tilde{\lambda}_1) \sinh(2 T \omega) + \nn \\
&&+ (\tilde{\lambda}_{-3} + \tilde{\lambda}_{-1} + \tilde{\lambda}_1 ) \sinh(4 T \omega)\Big]\\
\alpha_1(T)\;&=&\;-\frac{1}{192 \omega \sinh^4(T \omega)}\Big[4(3 \tilde{\lambda}_{-3} - \tilde{\lambda}_{-1} + 3 \tilde{\lambda}_1) T \omega \cosh(T \omega) + \nn \\
&&+ (11 \tilde{\lambda}_{-3} + \tilde{\lambda}_{-1} - 9 \tilde{\lambda}_1) \sinh(T \omega) + (3 \tilde{\lambda}_{-3} + \tilde{\lambda}_{-1} - \tilde{\lambda}_1) \sinh(3 T \omega)  \Big]\\
\alpha_2(T)\;&=&\frac{1}{192 \omega \sinh^4(T \omega)} \Big[4 (3 \tilde{\lambda}_{-3} - \tilde{\lambda}_{-1} + 3 \tilde{\lambda}_1) T \omega + 2 (3 \tilde{\lambda}_{-3} - \tilde{\lambda}_{-1} + 3 \tilde{\lambda}_1) T \omega \cosh(2 T \omega) + \nn \\
&&+ 3 (5 \tilde{\lambda}_{-3} + \tilde{\lambda}_{-1} - 3 \tilde{\lambda}_1) \sinh(2 T \omega) \Big]\;
\ea
These coefficients describe Hamilton's principal function corresponding to a perturbation term (as this is still first order in $\lambda$)
\ba
\frac{\lambda}{4!}\Big(\,\frac{\tilde{\lambda}_{-3}}{\omega^4} \dot{x}^4(t)\, +\, \frac{\tilde{\lambda}_{-1}}{\omega^2} \dot{x}^2(t) x^2(t) \,+\, \tilde{\lambda}_1 x^4(t) \Big)
\ea
added to the Lagrangian for the harmonic oscillator. The solutions that start with $n_0=-3$, $n_0=-1$ correspond to terms $\dot x^4$ and $\dot x^2x^2$ added to the Lagrangian.

Applying the same techniques to the last recursion equation (\ref{Gam}) reveals that there is only one free parameter $\xi$ for the solutions, which does appear in the solution (\ref{Gams}).\\[5pt]

To summarize, we have found all the solutions to the fixed point equations under the (physically justifiable) assumption that all the solutions can be expanded into a power series in $T$ starting with some lowest, not necessarily positive, order. The free parameters that do appear in the solutions can be all interpreted in terms of `large scale' physics. Hence there are no discretization ambiguities left, if we require reparametrization invariance to hold.

\section{Recursion relations for the $\beta_i$}\label{App:BetaRecursion}

Here we want to shortly explain how one can tackle the recursion relations (\ref{b32})
\ba\label{bd1}
 \beta_0^{(n+1)}(2T)\;&=&\;\beta_0^{(n)}\left(T\right)\left(1+\frac{1}{2
\cosh^2(T\omega)}\right)\;+\;\beta_1^{(n)}\left(T\right)\frac{1}{2
\cosh(T \omega)}\\[5pt]\nonumber
&&\;+\;
\frac{\hbar}{\omega} \Big( \frac{3 \tanh(T \omega)}{2 \cosh^2(T \omega)} \alpha_0^{(n)}(T) + \frac{3 \tanh(T \omega)}{4 \cosh(T \omega)} \alpha_1^{(n)}(T) + \frac{\tanh(T\omega)}{2} \alpha_2^{(n)}(T) \Big)
\\[5pt]\label{Gl:RecursionForBeta1a}
\beta_1^{(n+1)}(2T)\;&=&\;\beta_0^{(n)}\left(T\right)\frac{1}{\cosh^2(T\omega)}\;+\;\beta_1^{(n)}\left(T\right)\frac{1}{
\cosh(T\omega)}\\[5pt]\nonumber
&&\;+\;
\frac{\hbar}{\omega}\Big(\frac{3 \tanh(T \omega)}{\cosh^2(T \omega)} \alpha_0^{(n)}(T)+ \frac{3 \tanh(T \omega)}{2 \cosh(T\omega)} \alpha_1^{(n)}(T)\Big)
\ea
These recursion relations have  a homogeneous part
\ba\label{bd2}
\beta_0^{(n+1)}(2T)\;&=&\;\beta_0^{(n)}\left(T\right)\left(1+\frac{1}{2
\cosh^2(T\omega)}\right)\;+\;\beta_1^{(n)}\left(T\right)\frac{1}{2
\cosh(T \omega)}\\[5pt]\label{Gl:RecursionForBeta1}
\beta_1^{(n+1)}(2T)\;&=&\;\beta_0^{(n)}\left(T\right)\frac{1}{\cosh^2(T\omega)}\;+\;\beta_1^{(n)}\left(T\right)\frac{1}{
\cosh(T\omega)}    \q .
\ea
to which an inhomogeneity is added. Hence there will be at least one ambiguity, as to every solution of the inhomogeneous equations one can add an arbitrary multiple of the solutions to the homogeneous solutions. One therefore needs to determine the solutions to the homogeneous equations, if one wants to find the full space of solutions. These solutions can be obtained as explained in the main text, or alternatively, by subsequently transforming the variables to simplify and decouple the equations. Another general method is the expansion in a power series in $T$, which replaces the `coupling functions' $\beta_0(T)$, $\beta_1(T)$ by infinitely many coupling constants $\beta_0^k$, $\beta_1^k$.

Assume we are being given the solution $\beta^h_0$, $\beta^h_1$ to the homogeneous equations, in our case
\ba\label{bd3}
\beta_0^h(T)\;&=&\;\mu \coth(T\omega)\,-\,\nu \frac{T \omega}{\sinh^2(T\omega)} \nn\\
\beta_1^h(T)\;&=&\;\nu \frac{2T \omega \cosh(T)}{\sinh^2(T)}\,-\,\mu \frac{2}{\sinh(T)}
\ea

We want to introduce new variables $\tilde \beta_0$, $\tilde \beta_1$ such that the recursion relations (\ref{bd1}) decouple and simplify. To this end we can use the solutions (\ref{bd3}). To decouple the equations we choose the new variables such that $\tilde \beta_0 \sim \mu$ and $\tilde \beta_1 \sim \nu$ for the solutions in the new variables:
\ba
\tilde{\beta}_0(T)\;&:=&\;2 \cosh(T \omega ) \beta_0(T) + \beta_1(T) \nn\\
\tilde{\beta}_1(T)\;&:=&\; \beta_0(T) + \frac{1}{2} \cosh(T \omega) \beta_1(T)   \q .
\ea
Indeed we find for the  (homogeneous) fixed point conditions 
%
\ba
\tilde{\beta}^{*}_0(2T)\;&=&\; 2 \cosh(T \omega) \tilde{\beta}^{*}_0(T) 
\nn\\
\tilde{\beta}^{*}_1(2T) \;&=&\; 2 \tilde{\beta}^{*}_1(T) 
\q .
\ea

\noindent The homogeneous solutions in the new variables are now
\ba
\tilde{\beta}_0^{h}(T)\;&=&\; 2 \mu \sinh(T \omega) \\
\tilde{\beta}_1^{h}(T)\;&=&\; \nu T \omega \q ,
\ea
To write the fixed point equations into the form $f(2T) -f(T) =g(T)$ with $g(T)$ representing the inhomogeneous terms, we apply another transformation:
\ba
\bar{\beta}_0(T)\;&:=&\;\frac{\tilde{\beta}_0(T)}{\sinh(T \omega)} \\
\bar{\beta}_1(T)\;&:=&\;\frac{\tilde{\beta}_1(T)}{T} \q .
\ea
%
 This will lead to the following inhomogeneous fixed point conditions (where we replaced the $\alpha^{(n)}_k$ by the fixed point (\ref{b31a}))
\ba
\bar{\beta}^{*}_0(2T)- \bar{\beta}^{*}_0(T) &=&
  \frac{\hbar}{32 \omega^2 \sinh^4(2 T \omega)} \Big[ \cosh(T \omega) ( 10 T \omega + 2 T \omega \cosh(4 T \omega) - 4 \sinh(2 T \omega) - \nn\\ &&\q\q\q\q \q \sinh(4 T \omega)) +
2 ( - 6 T \omega \cosh(2 T \omega) + \sinh(2 T \omega) + \sinh(4 T \omega))\Big] \nn\\
\bar{\beta}^{*}_1(2T) -\bar{\beta}^{*}_1(T)  &=&
 \frac{\hbar \sinh^2(\frac{T \omega}{2})}{32 T \omega^2 \sinh^3 (2 T \omega)} \Big[ -8 T \omega -  6 T \omega \cosh(T \omega) -  2 T \omega \cosh(2 T \omega) + \nn\\
 &&\q\q\q\q 2 T \omega \cosh(3 T \omega)
+ \sinh(T \omega) + 3 \sinh(2 T \omega) + \sinh(3 T \omega) + \sinh(4 T\omega)\Big] \nn\\
\ea
The fixed point condition for both recursion relations is now of the form
$f(2T)-f(T)=g(T)$
where $g(T)$ is some known function of $T$. For this one can use an `integration table', see below. Alternatively a power series expansion of the solution can be easily given: If $g=\sum_{n\neq 0} g_n T^n$ then
\ba
f(T)=\sum_{n\neq 0} \frac{g_n}{2^n-1} T^n  \q .
\ea
Note that it would be inconsistent to have a constant term $\sim T^0$ in $g(T)$. An arbitrary integration constant can be added to the solution $f(T)$. This ambiguity in the solutions corresponds however to adding an arbitrary multiple of the homogeneous solutions (which here are the constants) to one solution of the inhomogeneous recursion relations.

The following table can be used to obtain a more explicit solution for $f(T)$.
\begin{eqnarray}\label{Tab:Table}
\begin{array}{|c|c|}
\hline
\\\q\q \q\q f(T) \qquad\qquad&\qquad\qquad g(T)=f(2T)\,-\,f(T) \q\q\q\q \\[10pt] \hline
T & T\\[5pt] \hline
\coth^2T & -\frac{\cosh 2T}{\sinh^2 2T}\,-\,\frac{1}{2\sinh^2T}\\[5pt]\hline
\coth T & \frac{1}{\sinh 2T}\\[5pt] \hline
%
%
\frac{T}{\sinh^2T} & -2T\,\frac{\coth 2T}{\sinh 2T}\\[5pt] \hline
\tanh T & \frac{\tanh T}{\cosh 2T}\\[5pt] \hline
T \coth T & T\tanh T\\[5pt] \hline
\frac{T}{\sinh 2T} & 2T\frac{1-\cosh T}{\sinh 2T}\\[5pt] \hline
T \tanh T & T\tanh 2T\,+\,T\frac{\tanh T}{\cosh 2T}\\[5pt] \hline
\ln\sinh T & \ln(2\cosh T)\\[5pt] \hline
-\frac{2}{T}\coth T& \frac{1}{T}\left(\frac{1}{\sinh(2T)}+\coth(T)\right)\\[5pt] \hline
T f'(T) & T g'(T) \\[5pt]
\hline
\end{array}
\end{eqnarray}

\noindent In this way one can find all solutions to the fixed point equations for the $\beta_i(T)$:
\ba
\beta_0^{*}(T)\;&=&\;\mu\coth(T \omega)\,-\,\nu T\omega \frac{1}{\sinh^2(T\omega)} 
%
\,+\,\frac{\tilde \lambda\hbar}{32 \omega^2 \sinh^2(T \omega)}\Big(2 + \cosh^2(T \omega) - 3 T \omega \coth(T \omega)\Big) \nn\\
\beta_1^{*}(T)\;&=&\;\nu\frac{2T\omega \cosh(T \omega)}{\sinh^2(T \omega)}\,-\,\mu
\frac{2}{\omega^2 \sinh(T \omega)}\,+ \nn\\
&&
\frac{\tilde \lambda\hbar}{32 \sinh^3(T \omega)}\Big(4 T \omega + 2 T \omega \cosh(2 T \omega) - 3 \sinh(2 T\omega)\Big)  \q .
\ea

\end{document}